\documentclass[prodmode]{acmsmall} 

\usepackage[ruled]{algorithm2e}

\SetAlFnt{\small}
\SetAlCapFnt{\small}
\SetAlCapNameFnt{\small}
\SetAlCapHSkip{0pt}
\IncMargin{-\parindent}
\usepackage[numbers]{natbib}

\newif\ifFULL
\FULLtrue

\usepackage{complexity}

\newcommand{\ignore}[1]{}

\usepackage{nameref}
\usepackage{bbm}

\usepackage{thmtools}
\usepackage{thm-restate}
\ignore{
\declaretheorem[name=Theorem,sibling=thm]{theom}
\declaretheorem[name=Lemma,sibling=lem]{againlem}
\declaretheorem[name=Corollary,sibling=cor]{againcor}
\declaretheorem[name=Proposition,sibling=prop]{againprop}
}

\makeatletter
\let\orgdescriptionlabel\descriptionlabel
\renewcommand*{\descriptionlabel}[1]{%
  \let\orglabel\label
  \let\label\@gobble
  \phantomsection
  \edef\@currentlabel{#1}%
  \let\label\orglabel
  \orgdescriptionlabel{#1}%
}
\makeatother
\newtheorem{informal}{Informal Theorem}
\newcommand{\srev}{\textsc{SRev}}
\newcommand{\val}{\textsc{Val}}
\newcommand{\brev}{\textsc{BRev}}
\newcommand{\rev}{\textsc{Rev}}
\newcommand{\srevstar}{\textsc{SRev}^*}

\newdef{defn}[theorem]{Definition}

\usepackage{mathtools}
\usepackage{amsmath,amssymb,amsfonts}
\ignore{
\usepackage{latexsym,bbm,xspace,graphicx,float,mathtools}
\usepackage[backref,colorlinks,citecolor=blue,bookmarks=true]{hyperref}
\usepackage[letterpaper,margin=1in]{geometry}
\usepackage{color}
\usepackage{algorithmic}

\usepackage{geometry} 
\geometry{a4paper} 

\usepackage{graphicx} 

\ignore{
\usepackage{booktabs} 
\usepackage{array} 
\usepackage{paralist} 
\usepackage{verbatim} 
\usepackage{subfig} 
}
\usepackage{fancyhdr} 
\pagestyle{fancy} 
\lhead{}\chead{}\rhead{}
\lfoot{}\cfoot{\thepage}\rfoot{}

\ignore{
\usepackage[nottoc,notlof,notlot]{tocbibind} 
\usepackage[titles,subfigure]{tocloft} 

}

\usepackage{color}

  \providecommand{\lemmaname}{Lemma}
  \providecommand{\propositionname}{Proposition}
\providecommand{\corollaryname}{Corollary}
\providecommand{\theoremname}{Theorem}
}

\begin{document}

\markboth{Rubinstein and Weinberg}{Simple Mechanisms for a Subadditive Buyer and Applications to Revenue Monotonicity}

\title{Simple Mechanisms for a Subadditive Buyer\\ and Applications to
Revenue Monotonicity}
\author{
AVIAD RUBINSTEIN$^1$
\affil{UC Berkeley}
S. MATTHEW WEINBERG$^2$ 
\affil{Princeton University}
}

\begin{abstract}
We study the revenue maximization problem of a seller with $n$ heterogeneous
items for sale to a single buyer whose valuation function for sets
of items is unknown and drawn from some distribution $D$. We show
that if $D$ is a distribution over subadditive valuations with independent
items, then the better of pricing each item separately or pricing
only the grand bundle achieves a constant-factor approximation to
the revenue of the optimal mechanism. This includes buyers who are
$k$-demand, additive up to a matroid constraint, or additive up to
constraints of any downwards-closed set system (and whose values for
the individual items are sampled independently), as well as buyers
who are fractionally subadditive with item multipliers drawn independently.
Our proof makes use of the core-tail decomposition framework developed
in prior work showing similar results for the significantly simpler
class of additive buyers~\cite{LiY13, BabaioffILW14}.

In the second part of the paper, we develop a connection between approximately
optimal simple mechanisms and approximate revenue monotonicity with
respect to buyers' valuations. Revenue non-monotonicity is the phenomenon
that sometimes strictly \emph{increasing}  buyers' values for every
set can strictly \emph{decrease}  the revenue of the optimal mechanism~\cite{HartR12}.
Using our main result, we derive a bound on how bad this degradation
can be (and dub such a bound a proof of \emph{approximate}  revenue
monotonicity); we further show that better bounds on approximate monotonicity
imply a better analysis of our simple mechanisms.
\end{abstract}

\ccsdesc[500]{Theory of computation~Algorithmic game theory and mechanism design}
\ccsdesc[500]{Theory of computation~Algorithmic mechanism design}
\ccsdesc[500]{Theory of computation~Computational pricing and auctions}

\keywords{Revenue optimization, combinatorial valuations, simple auctions, revenue monotonicity}


\begin{bottomstuff}
$^1$ Department of Electrical Engineering and Computer Science, UC Berkeley, aviad@eecs.berkeley.edu.
This research was supported by NSF grants CCF0964033 and CCF1408635, and by Templeton Foundation grant 3966. 
This work was done in part at the Simons Institute for the Theory of Computing and while visiting Princeton University.

$^2$ Department of Computer Science, Princeton University, smweinberg@princeton.edu. 
\end{bottomstuff}

\maketitle


\section{Introduction}

Consider a revenue-maximizing seller with $n$ heterogeneous items
for sale to a single buyer whose value for sets of items is unknown,
but drawn from a known distribution $D$. When $n=1$, seminal work
of Myerson~\cite{Myerson81} and Riley and Zeckhauser~\cite{RileyZ83}
shows that the optimal selling scheme simply sets the price $p^{*}=\arg\max\{p\cdot\Pr[v\geq p|v\leftarrow D]\}$.
Thirty years later, understanding the structure of the optimal mechanism
when $n>1$ still remains a central open problem. Unfortunately, it
is well-known that the optimal mechanism may require randomization,
behave non-monotonically, and be computationally hard to find, even
in very simple instances~\cite{Thanassoulis04, Pavlov11, BriestCKW10, DaskalakisDT14, ChenDPSY14, HartN13, HartR12, PPPR16-dynamic}.
In light of this, recent work began studying the performance of especially
simple auctions through the lens of approximation. Remarkably, these
works have shown that when the bidder's valuation is \emph{additive}%
\footnote{A valuation function $v(\cdot)$ is additive if $v(S\cup T)=v(S)+v(T)$
for all $S\cap T=\emptyset$.%
}, and her value for each item is drawn independently, very simple
mechanisms can achieve quite good approximation ratios. Specifically,
techniques developed in this series of works proves that the better
of setting Myerson's reserve on each item separately or setting Myerson's
reserve on the grand bundle of all items together achieves a $6$-approximation~\cite{HartN12, LiY13, BabaioffILW14}. 

While this model of buyer values is certainly mathematically interesting
and economically motivated, it is also perhaps too simplistic to have
broad real-world applications. A central question left open by these
works is whether or not simple mechanisms can still approximate optimal
ones in more general settings. In this work we resolve this question
in the affirmative, showing that the better of selling separately
(we will henceforth use $\textsc{SRev}$ to denote the revenue of
the optimal such mechanism) or together (henceforth \brev) still
obtains a constant-factor approximation to the optimal revenue (henceforth
\rev) when buyer values are combinatorial in nature but complement-free. 

\begin{informal}\label{inf:main}
Let $D$ be any distribution over subadditive valuation functions with independent items. Then $\max\{\textsc{SRev},\textsc{BRev}\}\geq \Omega(1)\cdot \textsc{Rev}$. Furthermore, prices providing this guarantee can be found computationally efficiently. 
\end{informal}

We postpone a formal definition of exactly what it means for $D$
to have ``independent items'' to Section \ref{sec:Prelim}. We note
here a few instantiations of our model in commonly studied settings
(from least to most general):
\begin{itemize}
\item $k$-demand: The buyer has value $v_{i}$ for item $i$, and the $v_{i}$s
are drawn independently. The buyer's value for a set $S$ is $v(S)=\max_{T\subseteq S,|T|\leq k}\{\sum_{i\in T}v_{i}\}$. 
\item Additive up to constraints $\mathcal{I}$: $\mathcal{I}$ is some
downwards-closed set system on $[n]$. The buyer has value $v_{i}$
for item $i$, and the $v_{i}$s are drawn independently. $v(S)=\max_{T\subseteq S,T\in\mathcal{I}}\{\sum_{i\in T}v_{i}\}$.
\item Fractionally-subadditive: buyer has ``possible values'' $\{v_{ij}\}_{j}$
for item $i$, and the sets $\{v_{ij}\}_{j}$ are drawn independently
across items (but may be correlated within an item). $v(S)=\max_{j}\{\sum_{i\in S}v_{ij}\}$. 
\end{itemize}

A recent book of Hartline~\cite{hartlinebook} provides a fantastic discussion of the role of approximation in mechanism design. Before proceeding, it is worth repeating some aspects of this discussion to view our result in the proper context. One should not interpret our main result as claiming that sellers should be satisfied with a constant fraction of the optimal obtainable revenue, but rather as studying the tradeoff between simplicity and optimality. Sometimes, the optimal mechanism simply isn't an option: perhaps it is prohibitively complex to implement, prohibitively frustrating for buyers to participate, or prohibitively difficult (computationally) to find. And even when the optimal mechanism is a feasible option, the desire for simplicity and transparency may outweigh the expected loss in revenue. Similarly, one should not interpret the ratios obtained in our main result (they are noticeably larger than $6$) as ratios that one might expect to trade off in practice, as these are \emph{provable} bounds for \emph{worst-case} instances.

\subsection{Challenges of Combinatorial Valuations}

The design of simple, approximately optimal mechanisms for any non-trivial
multi-item setting has been a large focus for much of the Algorithmic
Game Theory community over the past decade. Even ``simple'' settings
with additive or unit-demand valuations required significant breakthroughs.
The key insight enabling these breakthroughs for additive buyers is
that the buyer's valuation is separable across items. While the optimal
mechanism can still be quite bizarre despite this realization~\cite{HartR12},
this fact enables certain elementary decomposition theorems that are
surprisingly powerful (e.g. the ``Marginal Mechanism''~\cite{CaiH13,HartN12}).
However, these theorems are extremely sensitive to being able to separate
the marginal contribution of different items \emph{exactly} (and
not just via upper/lower bounds). This is due to the phenomenon that
a slight miscalculation in estimating a buyer's value may cause her
to change preferences entirely, resulting in a potentially unbounded
loss of revenue. One of our main technical contributions is overcoming
this obstacle by providing an \emph{approximate} version of these
decomposition theorems.

A further complication in applying these previous techniques is that
they all make use of the fact that $\textsc{SRev}(D_1 \times \ldots \times D_n) = \sum_i \textsc{SRev}(D_i)$.
This claim is not even approximately true for subadditive buyers,
and the ratio between the two values could be as large as $n$ (the
right-hand side is always larger). To have any hope of applying these
tools, we therefore need a proxy for \textsc{SRev} that at least
approximately has this separability property.

For unit-demand buyers, the key insight behind the mechanisms designed
in~\cite{ChawlaHK07, ChawlaHMS10, CMS10_eps-BIC, KleinbergW12} is that
every multi-dimensional problem instance has a related single-dimensional
problem instance, and there is a correspondence between truthful mechanisms
in the two instances. This realization means that one can instead
design mechanisms for the single-dimensional setting, where optimal
mechanisms are well understood due to Myerson's virtual values, and
translate them in a black-box manner to mechanisms for the original
instance. While these techniques have proven extremely fruitful in
the design of mechanisms for multiple unit-demand buyers and sophisticated
feasibility constraints, they have also proven to be limited in use
to unit-demand settings. A special case of our results can be seen
as providing an alternative proof of the single-buyer result of Chawla,
Hartline, and Kleinberg~\cite{ChawlaHK07} (albeit with a significantly
worse constant) that doesn't require virtual valuation machinery.

Aside from the difficulties in applying existing machinery to design
optimal mechanisms for combinatorial valuations, formal barriers exist
as well. For instance, it is a trivial procedure for an additive buyer
to select his utility-maximizing set of items when facing an item-pricing,
and finding the revenue-optimal item-pricing is also trivial (just
find the optimal price for each item separately). Yet for a subadditive
buyer, both tasks are quite non-trivial. Just \emph{computing} the
expected revenue obtained by a fixed item-pricing is \NP-hard. Worse,
the buyer's problem of just \emph{selecting}  her utility-maximizing
set from a given item-pricing is also \NP-hard! Therefore, buyers
may behave quite unpredictably in the face of an item-pricing depending
on how well they can optimize. Moreover, even if we are willing to
assume that the buyer has the computational power to select her utility-maximizing
set, it is known \emph{still}  that (without our independence assumption)
finding an $n^{c}$-approximately optimal mechanism is \NP-hard for
all $c=O(1)$~\cite{CaiDW13b}. We sidestep all these difficulties
by not attempting to compute or approximate $\textsc{SRev}$ at all,
nor trying to predict bizarre buyer behavior. We instead perform our
analysis on revenue contributions only of items purchased \emph{when the buyer is not willing to purchase any others}.
Buyer behavior in such instances is predictable and easily computable:
simply purchase the unique item for which $v(\{i\})>p_{i}$. It is
surprising that such an analysis suffices, as it completely ignores
any revenue contribution coming from the entirely plausible event
that the buyer is willing to purchase multiple items.


\subsection{Techniques}

We prove our main theorem by making use of the core-tail decomposition
framework introduced by Li and Yao~\cite{LiY13}. There are three
high-level steps to applying the framework. The first is proving a
``core decomposition'' lemma that separates the optimal revenue into
contributions from items which the buyer values very highly (the ``tail''),
and items which the buyer values not so high (the ``core''). The
second is showing that the contribution from the tail can be approximated
well by \srev. The third is showing that the contribution from the
core can be approximated well by $\max\{\srev,\brev\}$.

\paragraph{The Core Decomposition Lemma}

The proof of the original Core Decomposition Lemma in~\cite{LiY13}
was obtained by cleverly stringing together simple claims proved in~\cite{HartN12}.
As discussed above, these seemingly ``obvious'' claims may not extend
beyond additive valuations over independent items, due to the fact
that the buyer's value cannot be separated across items. Nevertheless,
we are able to prove an approximate version of the core decomposition
lemma for subadditive buyers (Lemma \ref{lem:(Approximate-Core-Decomposition)})
by making use of ideas from reductions from $\epsilon$-truthful mechanisms
to fully truthful ones. Like in~\cite{BabaioffILW14}, our core decomposition
lemma holds for many buyers. The proof for a single buyer, which is
the focus of this paper can be found in Section \ref{sub:cdl}.
\ifFULL
We also provide (Section~\ref{app:many}) a more technically involved
proof for many buyers which builds on heavier tools from~\cite{BeiH11, HartlineKM11, DaskalakisW12}.
\fi

\paragraph{Bounding the Tail's Contribution}

Arguments for bounding the contribution from the tail in prior work
(and ours) use the following reasoning. If the cutoff between core
and tail is sufficiently high, then the probability that $k$ items
are simultaneously in the tail for a sampled valuation decays exponentially
in $k$. If one can also show that the approximation guarantee of
$\textsc{SRev}$ decays subexponentially in $k$, then we can bound
the gap between $\textsc{SRev}$ and the tail's contribution by a
constant factor. We show that indeed the approximation guarantee of
$\textsc{SRev}$ decays only polynomially in $k$.

\paragraph{Bounding the Core's Contribution}

Arguments for bounding the contribution from the core in prior work
(and ours) use the following reasoning. The total expected value for
items in the core is a subadditive function of independent random
variables (bounded above by the core-tail cutoff). If the cutoff between
core and tail is sufficiently low, then one of two things must happen.
Either the expected contribution from the core is also small, in which
case $\textsc{SRev}$ itself provides a good approximation, or the
expected contribution is large, and therefore also large with respect
to the cutoffs. In the latter case, a concentration bound implies
that $\textsc{BRev}$ must provide a good approximation. In the additive
case, the appropriate concentration bound is Chebyshev's inequality.
In the subadditive case, we need heavier tools, and apply a concentration
bound due to Schechtman \cite{Sch99-concentration_results}.

\subsection{Connection to Approximate Revenue-Monotonicity}

Consider designing revenue-optimal mechanisms for two different markets,
and suppose that the valuations of the consumers in the first market
first-order stochastically dominate%
\footnote{We say that a distribution $D^{+}$ over valuation functions $v^{+}$
first-order stochastically dominates distribution $D$ over valuation
functions $v$ if the probability spaces can be coupled so that for
every subset $S$, $v^{+}\left(S\right)\geq v\left(S\right)$.%
} the valuations of the consumers in the second market.  It then seems
reasonable to expect that the optimal revenue achieved from the first
market, $\rev(D^{+})$, should be at least as large as the revenue
achieved from the dominated market, $\rev(D)$. When there is just
a single item for sale, this is an easy corollary of the format for
Myerson's optimal auction. Yet Hart and Reny provided an example where
this intuition breaks even in a setting as simple as an \emph{additive buyer with i.i.d. values for two items}~\cite{HartR12}.
Surprisingly, their example shows that it is possible to make \emph{strictly}
more revenue in a market when buyers have \emph{strictly} less value
for your goods, and the market need not even be very complex for this
phenomenon to occur.

A natural question to ask then, is how large this anomaly can be.
For example, Hart and Reny's constructions exhibit a (multiplicative)
gap of $33/32$ between $\rev(D^{+})$ and $\rev(D)$ for an additive
buyer with correlated values for two items, and $(1+\frac{1}{7000000})$
for an additive buyer with i.i.d. values for two items. Interestingly,
the simple mechanisms of~\cite{HartN12, LiY13, BabaioffILW14} upper
bound the possible gap of any instance where their results apply,
since $\srev$ and $\brev$ are monotone for additive buyers (i.e.
$\srev(D^{+})\geq\srev(D)$ and $\brev(D^{+})\geq\brev(D)$). Specifically,
for an additive buyer the gap is at most $(1+1/e)$ for two i.i.d.
items, $2$ for two asymmetric independent items, and $6$ for any
number of independent items. In Section \ref{sec:monotonicity} we
show that as a corollary of our results, the gap is also constant
for a subadditive buyer with independent items. Interestingly, this
connection between approximately optimal simple mechanisms and approximate
revenue-monotonicity is also fruitful in the other direction: it turns
out that improving the bound on approximate monotonicity for a subadditive
buyer would also improve the constant in our main theorem. Finally,
we show in Section \ref{sub:Correlated-distributions-are-nonmonotone}
that for an additive buyer with correlated values for two items, the
gap is potentially infinite. (This is the case for which Hart and
Reny provide a gap of $33/32$.) The proof is by a black-box reduction
from an example due to Hart and Nisan \cite{HartN13} that exhibits
a similar gap between simple and optimal mechanisms\textcolor{black}{,
further demonstrating the connection between these two important research
directions.}

\subsection{Discussion, Related Work and Open Problems}

Our work contributes to the recent growing literature on simple, approximately
optimal mechanisms. We extend greatly beyond prior work, providing
the first simple and approximately optimal mechanisms for buyers with
combinatorial valuations. Prior to our work, virtually nothing was
known about this setting (modulo the impossibility result of~\cite{CaiDW13b}).
Our results also demonstrate the strength of the core-tail decomposition
framework developed by Li and Yao to go beyond additive buyers.

In our opinion, the most exciting open question in this area is extending
these results to multiple buyers. A beautiful lookahead reduction
was previously developed by Yao~\cite{Yao15} for additive buyers, but applying his techniques for combinatorial bidders proved quite challenging. Following a preliminary version of this paper, others have built upon our techniques and made quite significant progress on this problem~\cite{CDW16-duality, CM16-ex_ante, CZ16-XOS}. To put them in context with respect to ours:
\begin{itemize}
\item\cite{CDW16-duality} provide an alternative proof of Yao's result and improve the constant factor (from 69 to 8). They also provide a duality framework, and show that essentialy all bounds on the optimal revenue used in prior works can be derived by essentially the same dual solution. The only exception we are aware of is our ``approximate marginal mechanism'' lemma, which to date has no ``duality proof'' in their framework. It would be interesting to understand this lemma in terms of duality, but not essential for its application.
\item \cite{CM16-ex_ante} provide the first approximately optimal mechanism for multiple combinatorial buyers. Without getting into a full definition, their results apply to a special case of valuations that are ``gross substitutes over independent items.'' They also improve the constant factor in our analysis. Their work did make use of part of our analysis, but required substantial conceptual novelty.
\item Most recently,~\cite{CZ16-XOS} extended their result to multiple buyers that are ``XOS over independent items.'' Their result draws on numerous previous works, those referenced above~\cite{Yao15,CDW16-duality, CM16-ex_ante}, ours, and also the posted-price mechanism of~\cite{FeldmanGL15} for \emph{welfare-maximization} in Bayesian settings with multiple XOS bidders. They obtain a logarithmic approximation for multiple buyers that are subadditive over independent items, and the main bottleneck in extending their constant-factor approximation from XOS to subadditive is that Feldman et al.'s welfare guarantee only holds for XOS valuations. Obtaining a constant-factor approximation for many buyers with subadditve valuations remains a key open question, although each of the aforementioned papers makes \emph{substantial} progress beyond ours. 
\end{itemize}

Another important direction is
extending our understanding of simple mechanisms to models of limited
correlation over values for disjoint sets of items%
\footnote{Note that as we mention in the previous section, for arbitrary correlated
items the gap can be infinite \cite{HartN13,BriestCKW10}.%
}. Recent independent work of Bateni et. al.~\cite{BateniDHS15} addresses
this direction, providing approximation guarantees on $\max\{\srev,\brev\}$
vs. $\rev$ for an additive buyer whose values for items are drawn
from a common-base-value distribution and various extensions. Their results also make use of a core-tail decomposition, but the tools they develop beyond the decomposition are disjoint from ours. A natural
question in this direction is whether our results extend to settings
where buyer values are both combinatorial \emph{and} exhibit limited
correlation between disjoint sets of items, as the end goal is to
have a model that encompasses as many real-world instances as possible. 

\ignore{Very recent work of Cai et. al.~\cite{CaiDW16} provides a unifying primal-dual proof of many of the works that ours builds upon~\cite{ChawlaHK07, ChawlaHMS10, HartN12, LiY13, BabaioffILW14, Yao15}. Currently, however, there is no proof of our result in their framework. Finding a primal-dual proof of our result using their techniques is a challenging but promising direction, as such a proof would likely guarantee a significantly better approximation ratio and also be more amenable to multi-bidder extensions.}

Other related works include an approximately optimal mechanism for an additive buyer when the seller incurs production costs \cite{MS15-production_costs}, and an approximately optimal mechanism for a buyer with restricted complementarities \cite{EFFTW16-complements}. \cite{Rub16-optimal_simple} considers a related algorithmic problem of finding the optimal partition mechanism (where the seller posts a take-it-or-leave-it price on disjoint bundles). It remains unknown whether the optimal partition mechanism can provide provably better guarantees than $\max\{\srev,\brev\}$ with respect to $\rev$ in any domain. 

If we were to idenfity the contributions of the present work that seem to have been most useful in follow-up works, it would be:
\begin{itemize}
\item The ``approximate marginal mechanism lemma,'' Lemma~\ref{lem:(Approximate-Marginal-Mechanism)} and~Lemma~\ref{lem:marginalMany}, have been used (e.g. by Cai and Zhao) to reduce from subadditive/gross substitutes/etc. over independent items to ``additive subject to downwards-closed/matroid/etc. constraints with independent items'' at the cost of a constant factor.
\item Our application of Schechtman's bound in the core, Corollary~\ref{cor:concentration}, has been used in essentially every follow-up work discussing optimal mechanisms for combinatorial buyers~\cite{CM16-ex_ante,CZ16-XOS,EFFTW16-complements}.
\item In the face of buyers for whom selecting a utility-maximizing bundle might be NP-hard, the idea to count revenue \emph{only} when it is trivial to select a utility-maximizing bundle (e.g. because it is a singleton set) has proven useful in follow-up works such as~\cite{EFFTW16-complements}. 
\item Moreover, just a meaningful definition of ``subadditive over independent items'' seemed to be useful in all of the referenced works, and even for the design of combinatorial prophet inequalities (where again some notion of independent is necessary to subvert horrible lower bounds even in the additive case)~\cite{RS17-combinatorial_prophet}. 
\end{itemize}


\section{Preliminaries}
\label{sec:Prelim}

\ifFULL
We focus the initial exposition on the single-buyer problem, and
postpone all details regarding auctions for multiple buyers to Section~\ref{app:many}.
\fi
There is a single revenue-maximizing seller with $n$ items for sale
to a single buyer. The buyer has combinatorial valuations for the
items (i.e. value $v(S)$ for receiving set $S$), and is quasi-linear
and risk-neutral. That is, the buyer's utility for a randomized outcome
that awards him set $S$ with probability $A\left(S\right)$ while
paying (expected) price $p$ is $\sum_{S}A\left(S\right)v(S)-p$.
The valuation $v(\cdot)$ is unknown to the seller, who has a prior $D$ over possible
buyer valuations that is \emph{subadditive over independent items},
a term we describe below. By the taxation principle, the seller may
restrict attention to only lottery systems. In other words, the seller
provides a list of potential lotteries (distributions over sets) each
with a price, and the buyer chooses the utility-maximizing option.

\subsection{Subadditive valuations over independent items}

We now carefully define what we mean by subadditive valuations over
independent items. Intuitively, our model is such that the buyer has
some private information $x_{i}$ pertaining to item $i$,\footnote{Think of this information as ``information about the buyer's preferences related to item.''}
and $D^{\vec{x}}$ is a product distribution over $\mathbb{R}^{n}$
representing the seller's prior over the private information possessed
by the buyer. The buyer's valuation for set $S$ is parametrized by
the private information she has about items in that set, and can be
written as $V(\langle x_{i}\rangle_{i\in S},S)$. In economic terms, this models
that the items not received by the buyer impose no externalities.
We capture this formally in the definition below:
\begin{defn}
We say that a distribution $D$ over valuation functions $v(\cdot):\{ 0,1\} ^{n}\rightarrow\mathbb{R}$
is {\em subadditive over independent items} if:
\begin{enumerate}
\item All $v(\cdot)$ in the support of $D$ exhibit \emph{no externalities}.

Formally, let $\Omega_S = \bigtimes_{i\in S} \Omega_i$, where each $\Omega_i$ is a compact subset of a normed space. There exists a distribution $D^{\vec{x}}$ over $\Omega_{[n]}$
and functions $V_S:\Omega_S \rightarrow \mathbb{R}$ such that $D$ is the distribution that first
samples $\vec{x}\leftarrow D^{\vec{x}}$ and outputs the valuation
function $v(\cdot)$ with $v(S)=V_S(\langle x_{i}\rangle_{i\in S})$ for all
$S$.
\item All $v(\cdot)$ in the support of $D$ are \emph{monotone}. That
is, $v(S)\leq v(S\cup T)$ for all $S,T$.
\item All $v(\cdot)$ in the support of $D$ are \emph{subadditive}. That
is, $v(S\cup T)\leq v(S)+v(T)$.
\item The private information is \emph{independent across items}. That
is, the $D^{\vec{x}}$ guaranteed in property 1 is a product distribution.
\end{enumerate}
\end{defn}

We describe now how to encode commonly studied valuation distributions
in this model.
\begin{example}
The following types of distributions can be modeled as subadditive
over independent items. (Recall that $\vec{x}$
  is the vector of independently sampled attributes in the definition above.)
\begin{enumerate}
\item Additive: Let $\Omega_i = [0,1]$ and interpret $x_{i}$ as the buyer's value for item $i$. $V_S(\langle x_i\rangle_{i \in S})=\sum_{i\in S}x_{i}$.
\item $k$-demand: Let $\Omega_i = [0,1]$ and interpret $x_{i}$ as the buyer's value for item $i$. $V_S(\langle x_i\rangle_{i \in S})=\max_{T\subseteq S,|T|\leq k}\{\sum_{i\in T}x_{i}\}$.
\item Additive up to $\mathcal{I}$: Let $\Omega_i = [0,1]$ and interpret $x_{i}$ as the buyer's value for item
$i$. $V_S(\langle x_i\rangle_{i \in S})=\max_{T\subseteq S,T\in\mathcal{I}}\{\sum_{i\in T}x_{i}\}$.
\item Fractionally subadditive: Let $\Omega_i = [0,1]^k$ for any finite $k$ and interpret $x_{i}$ as encoding the values $\{v_{ij}\}_{j\in [k]}$. $V_S(\langle x_i\rangle_{i \in S})=\max_{j}\{\sum_{i\in S}v_{ij}\}$. 
\end{enumerate}
\end{example}

\subsection{Notation}
\begin{defn}
For any distribution $D$ of buyer's valuation, we use the following
notation, most of which is due to \cite{HartN12,BabaioffILW14}:
\begin{itemize}
\item $D_{i}$: The distribution of $v(\{i\})$ when $v(\cdot)\leftarrow D$.
\item $t$: the cutoff between core and tail. If $v(\{i\})>t$, we say that
item $i$ is in the tail. Otherwise it is in the core.
\item $D_{A}$: the distribution $D$, conditioned on $A$ being exactly
the set of items in the tail.
\item $D_{A}^{T}$: the distribution $D_{A}$ restricted just to items in
the tail (i.e. $A$). 
\item $D_{A}^{C}$: the distribution $D_{A}$ restricted just to items in
the core (i.e. $\bar{A}$). 
\item $p_{i}$: the probability that element $i$ is in the tail. 
\item $p_{A}$: the probability that $A$ is exactly the set of items in
the tail (note that $p_{\left\{ i\right\} }\neq p_{i}$).
\item $\textsc{Rev}\left(D\right)$: The maximum revenue obtainable via
a truthful mechanism from a buyer with valuation profile drawn from
$D$.
\item $\textsc{BRev}\left(D\right)$: The revenue obtainable from a buyer
with valuation profile drawn from $D$ by auctioning the grand bundle
via Myerson's optimal auction.
\item $\textsc{SRev}\left(D\right)$: The maximum revenue obtainable from
a buyer with valuation profile drawn from $D$ by pricing each item
separately. Note that when the buyer is not additive, $\srev(D)$
behaves erratically and is \NP-hard to find~\cite{ChenDPSY14}.
\item $\rev_{q}(D)$: For a one-dimensional distribution $D$, the optimal
revenue obtained by a reserve price that sells with probability at
most $q$. 
\item $\srevstar_{\vec{q}}(D)\ensuremath{:}\prod_{i}(1-q_{i})\cdot\sum_{i}\rev_{q_{i}}(D_{i})$:
a proxy for $\textsc{SRev}(D)$ that behaves nicer and is easy to
compute.
\item $\textsc{Val}\left(D\right)$: the buyer's expected valuation for
the grand bundle, $\E_{v\leftarrow D}\left[v\left(\left[n\right]\right)\right]$.
\end{itemize}
When the distribution is clear from the context, we simply use $\textsc{Rev}$,
$\textsc{Val}$, etc. Most of this notation is standard following~\cite{HartN12},
with the exception of $\rev_q$ and $\srevstar_{\vec{q}}$. We introduce
$\srevstar_{\vec{q}}$ because it will serve as a proxy to $\textsc{SRev}$
that behaves nicely and is easy to compute. Note that $\srevstar_{\vec{q}}$
is essentially computing the revenue of the best item pricing that
sells item $i$ with probability at most $q_{i}$, but only counting
revenue from cases where the other values are too low to have possibly
sold (and actually it undercounts this quantity). 
\end{defn}

\begin{remark}
In our definitions of $\rev_{q}(D)$ and $\srevstar_{\vec{q}}(D)$ we assume without loss of generality that for every single-dimensional $D$  and $q\in [0,1]$  it is possible to set a price that sells with probability exactly $q$.
When $D$ is a continuous distribution, this is true by the intermediate value theorem. 
When $D$ has a point mass, this is no longer true per se. 
Fortunately, there are standard methods for reducing the study of arbitrary distributions to continuous ones with arbitrarily small loss. We briefly sketch one, a rounding scheme commonly attributed to Nisan (that appears also in~\cite{ChawlaHK07,CaiH13}):

For any $\epsilon > 0$, $D$ can be ``smoothed'' into a continuous distribution $D^{\epsilon}$ by multiplying samples from $D$ by a random factor drawn uniformly from $[1,1+\epsilon]$. For any smoothed $D^{\epsilon}$, the desired prices exist by the intermediate value theorem. Using techniques similar in spirit to those of Section~\ref{sub:cdl}, it is easy (both computationally and conceptually) to convert mechanisms for $D^{\epsilon}$ to mechanisms for $D$, and vice versa, with negligible (dependent on $\epsilon$) loss in revenue. Therefore, one may formally study $D^\epsilon$ for sufficiently small $\epsilon$, and all results hold with respect to $D$ as well with negligible loss (and taking $\epsilon \rightarrow 0$ results in no loss at all). So for the remainder of the paper, we will assume w.l.o.g. that all distributions are continuous, and therefore the desired prices exist.

\end{remark}

We conclude the
preliminaries by stating a lemma of Hart and Nisan that we will use.
We include the proof below for completeness, as well as
to verify that it continues to hold when the valuations are not additive.

\begin{lemma} [Sub-domain Stitching special case~\cite{HartN12}]%
\label{lem:sds}
$\textsc{Rev}(D)\leq\sum_{A}p_{A}\textsc{Rev}(D_{A})$.
\end{lemma}
\begin{proof}
Let $M$ be an optimal mechanism for selling items with valuations
sampled from $D$, and let $\rev_{M}\left(D\right)=\rev\left(D\right)$
denote its revenue. Then, $\rev_{M}\left(D\right)=\sum p_{A}\rev_{M}\left(D_{A}\right)$.
Also, for each $A\subseteq\left[n\right]$, $\rev_{M}\left(D_{A}\right)\leq\rev\left(D_{A}\right)$. 
\end{proof}

\section{Main Result: Constant-factor approximation for subadditive buyer}
\begin{theorem}
\label{thm:main}When $D$ is subadditive over independent items,
there exists a probability vector $\vec{q}$ such that: 
\[
\rev(D)\leq314\srevstar_{\vec{q}}(D)+24\brev(D)
\]

Furthermore, $\vec{q}$ can be computed efficiently, as well as an
induced item pricing that yields expected revenue at least $\srevstar_{\vec{q}}(D)$. 
\end{theorem}

\subsubsection*{Proof outline}

We follow the core-tail decomposition framework. First, we provide
an approximate core decomposition lemma in Section \ref{sub:cdl}.
Then, we provide a bound on the contribution of the core with respect
to $\max\left\{ \brev,\srevstar\right\} $ in Section \ref{sub:Core},
and a bound on the contribution of the tail with respect to $\srevstar$
as a function of the cutoffs chosen in Section \ref{sub:Tail}. 

For ease of exposition, we simply set $t$ so that the probability
of having an empty tail is exactly half; i.e. $p_{\emptyset}=\prod\left(1-p_{i}\right)=1/2$.
We also set $\vec{q}=\vec{p}$.

\subsection{\label{sub:cdl}Approximate Core Decomposition}

In this section, we prove our approximate core decomposition lemma.
The key ingredient will be an approximate version of the ``Marginal
Mechanism'' lemma from \cite{CaiH13,HartN12} for subadditive buyers, stated
below:
\begin{lemma}
\label{lem:(Approximate-Marginal-Mechanism)}(``Approximate Marginal
Mechanism'') 

Let $S,X$ be a partition of $\left[n\right]$, and let $D=\left(D^{S},D^{X}\right)$
be the joint distribution for the valuations of items in $S,X$, respectively,
for buyers with subadditive valuations. Then for any $0<\epsilon<1$,
\textup{
\begin{equation}
\textsc{Rev}\left(D\right)\leq\left(\frac{1}{\epsilon}+\frac{1}{1-\epsilon}\right)\textsc{Val}\left(D^{S}\right)+\frac{1}{1-\epsilon}\E_{v^{S}\leftarrow D^{S}}\left[\textsc{Rev}\left(D^{X}\mid v^{S}\right)\right]\label{eq:marginal-mechanism}
\end{equation}
}

When $D^{S}$ and $D^{X}$ are independent, this simplifies to 
\[
\rev(D)\leq(\frac{1}{\epsilon}+\frac{1}{1-\epsilon})\val(D^{S})+\frac{1}{1-\epsilon}\rev(D^{X})\mbox{.}
\]

\end{lemma}
We outline the proof of Lemma \ref{lem:(Approximate-Marginal-Mechanism)}
below. We first recall the original ``Marginal
Mechanism'' lemma (that holds for an additive buyer without any multipliers).
We provide a complete proof so that the reader can see where the argument
fails for subadditive buyers. 
\begin{lemma}
\label{lem:HNmarginal}\emph{(``Marginal Mechanism''~\cite{CaiH13,HartN12})} Let
$S\sqcup X$ be any partition of $[n]$, and let $D^{+}$ be any distribution
over valuation functions such that $v^{+}(U)=v^{+}(U\cap S)+v^{+}(U\cap X)$
for all $U\subseteq n$, and $v^{+}$ in the support of $D^{+}$.
Let also $D^{S}$ denote $D^{+}$ restricted to items in $S$ and
$D^{X}$ denote $D^{+}$ restricted to items in $X$. Then $\rev(D^{+})\leq\val(D^{S})+\E_{v^{S}\leftarrow D^{S}}[\rev(D^{X}|v^{S})]$. \end{lemma}
\begin{proof}
We design a mechanism $M^{X}$ (the ``Marginal Mechanism'') to sell
items in $X$ to consumers sampled from $D^{X}|v^{S}$ based on the
optimal mechanism $M$ for selling items in $S\sqcup X$ to consumers
sampled from $D^{+}$. Define $A(v)$ to be the (possibly random)
allocation of items awarded to type $v$ in $M$, and $p(v)$ to be
the price paid. Let $M^{X}$ first sample a value $v^{S}\leftarrow D^{S}$.
The buyer is then invited to report any type $v^{X}$, and $M^{X}$
will award him the items in $A(v^{S},v^{X})\cap X$ and charge him
price $p(v^{S},v^{X})-v^{S}(A(v^{S},v^{X})\cap S)$. In other words,
the buyer will receive value from exactly the same items in $M^{X}$
as he would have received in $M$, except he receives the actual items
in $X$, whereas for items in $S$ he is given a monetary rebate instead
of his actual value. 

We first claim that if $M$ is truthful, then so is $M^{X}$. The
utility of a buyer with type $v^{X}$ for reporting $w^{X}$ to $M^{X}$
can be written as: $v^{X}(A(v^{S},w^{X})\cap X)+v^{S}(A(v^{S},w^{X})\cap S)-p(v^{S},w^{X})=(v^{S},v^{X})(A(v^{S},w^{X}))-p(v^{S},w^{X})$,
which is exactly the utility of a buyer with type $(v^{S},v^{X})$
for reporting $(v^{S},w^{X})$ to $M$. As $M$ was truthful, we
know that a buyer with type $(v^{S},v^{X})$ maximizes utility when
reporting $(v^{S},v^{X})$ over all possible $(v^{S},w^{X})$. Therefore,
a buyer with type $v^{X}$ prefers to tell the truth as well. 

Finally, we just have to compute the revenue of $M^{X}$. For each
$v^{S}$, the marginal mechanism provides a concrete mechanism for
the distribution $D^{X}|v^{S}$ that attains revenue at least $\textsc{Rev}(D^{+}|v^{S})-v^{S}(S)$.
So $\textsc{Rev}(D^{X}|v^{S})\geq\textsc{Rev}(D^{+}|v^{S})-v^{S}(S)$.
Taking an expectation over all $v^{S}$ (and an application of sub-domain
stitching) yields the lemma. 
\end{proof}
Notice that it is \emph{crucial} in the proof above that the buyer's
value could be written as $v^{S}(\cdot)+v^{X}(\cdot)$. Otherwise
the auctioneer does not know how much to ``reimburse'' the buyer,
since the correct amount depends on the buyer's private information.
The buyer can then manipulate his own report $v^{X}$ to influence
how much he gets reimbursed for the items in $S$. 

A natural approach then, given any distribution $D$ over subadditive
valuations, is to define a new value distribution $D^{+}$ by redefining
all $v(\cdot)$ to satisfy $v(U)=v(U\cap S)+v(U\cap X)$ (it is easy
to see that all valuations in the support of $D^{+}$ are still subadditive).
Unfortunately, even though $D^{+}$ (first-order stochastically) dominates
$D$, due to non-monotonicity we could very well have $\textsc{Rev}(D^{+})<\textsc{Rev}(D)$.
Still, we bound the revenue lost as we move from $D$ to $D^{+}$
by making use of tools for turning $\epsilon$-truthful mechanisms
into truly truthful ones. Lemma \ref{lem:delta} and Corollary~\ref{cor:eps-marginal-mechanism}
below capture this formally.

\begin{lemma}\label{lem:delta}Consider two coupled distributions $D$ and $D^{+}$,
with $v(\cdot)$ and $v^{+}(\cdot)$ denoting a random sample from
each. Define the random function $\delta(\cdot)$ so that $\delta(S)=v^{+}(S)-v(S)$
for all $S$. Suppose that $\delta(S)\geq0$ for all $S$ and that
$\delta(\cdot)$ also satisfies $\E_{D}\left[\max_{S\subseteq\left[n\right]}\{\delta\left(S\right)\}\right]\leq\overline{\delta}$.
Then for any $0<\epsilon<1$,

\[
\rev(D^{+})\geq(1-\epsilon)\left(\rev(D)-\bar{\delta}/\epsilon\right).
\]

\end{lemma}

\begin{proof}
Consider a mechanism $M$ which achieves the optimal revenue $\rev(D)$.
Let $(\phi_{v},p_{v})$ denote the lottery purchased by a buyer with
type $v$ in $M$, where $\phi_{v}$ is a (possibly randomized) allocation,
and $p$ is a price. Consider now the mechanism $M^+$ that offers
the same menu as $M$, but with all prices discounted by a factor
of $\left(1-\epsilon\right)$. Let $(\phi^+_{v},p^+_{v})$ denote the
lottery that a buyer with type $v^+$ (coupled with $v$) chooses to
purchase in $M^+$ (knowing that she would only pay $(1-\epsilon)p^+_{v}$ because of the discount). The following inequalities must hold (we will abuse
notation and let $v(\psi)=\E_{S\leftarrow \psi}[v(S)]$): 
\begin{eqnarray}
v(\phi_{v})-p_{v} & \geq & v(\phi^+_{v})-p^+_{v}\mbox{.}\label{eq:IC-original_buyer}\\
v^+(\phi^+_{v})-\left(1-\epsilon\right)p^+_{v} & \geq & v^+(\phi_{v})-\left(1-\epsilon\right)p_v\label{eq:IC-modified_buyer}
\end{eqnarray}
 Now, summing equations (\ref{eq:IC-original_buyer}) and (\ref{eq:IC-modified_buyer})
(then making use of the definition of $\delta(\cdot)$ and the fact
that it is non-negative), we have:
\begin{eqnarray*}
\epsilon p^+_{v}+\delta(\phi^+_{v}) & \geq & \epsilon p_v\\
\Rightarrow p^+_v & \geq & p_v-\delta(\phi^+_{v})/\epsilon
\end{eqnarray*}
We can now bound the expected revenue by taking an expectation over
all valuations:
\begin{eqnarray*}
\textsc{Rev}\left(D^+\right) & \geq & \E_{v\leftarrow D}\left[(1-\epsilon)p^+_{v}\right]\\
 & \geq & \left(1-\epsilon\right)\E_{v\leftarrow D}\left[p_{v}-\delta(\phi^+_{v})/\epsilon\right]\\
 & \geq & \left(1-\epsilon\right)\textsc{Rev}(D)-(1-\epsilon)\bar{\delta}/\epsilon
\end{eqnarray*}
\end{proof}

\begin{corollary}\label{cor:eps-marginal-mechanism}For a given partition of $[n]$,
$S\sqcup X$, and distribution $D$ over subadditive valuations, define
$D_{S}$ to be $D$ restricted to items in $S$, and $D^{+}$ to first
sample $v\leftarrow D$, and output $v^{+}(\cdot)$ with $v^{+}(U)=v(U\cap S)+v(U\cap X)$.
Then for all $\epsilon\in(0,1)$, $\rev(D)\leq\frac{\rev(D^{+})}{1-\epsilon}+\frac{\val(D_{S})}{\epsilon}$. 

\end{corollary}

\begin{proof}
By monotonicity, $v(U)\geq v(U\cap X)$ for all $U,X$. Therefore,
$v^+(U)-v(U)\leq v(U\cap S)\leq v(S)$ for all $U$. Furthermore, by
subadditivity, we have $v^+(U)\geq v(U)$ for all $U$. Together, this
means that $D$ and $D^+$ are coupled so that we can set $\delta(U)\leq v(S)$
for all $U$. Therefore, we may set $\bar{\delta}=\val(D_{S})$ in
the hypothesis of Lemma \ref{lem:delta}. The corollary follows by
rearranging the inequality.
\end{proof}

The proof of Lemma \ref{lem:(Approximate-Marginal-Mechanism)} is
now a combination of Corollary \ref{cor:eps-marginal-mechanism} and
Lemma \ref{lem:HNmarginal}.

\begin{proof}[of Lemma~\ref{lem:(Approximate-Marginal-Mechanism)}]
Chaining Corollary~\ref{cor:eps-marginal-mechanism} together with Lemma~\ref{lem:HNmarginal} we get:

$$\rev(D) \leq \frac{\rev(D^+)}{1-\epsilon} + \frac{\val(D^S)}{\epsilon}$$
$$\leq \frac{\val(D^S)+\E_{v^S \leftarrow D^S}\left[\rev(D^X|v^S)\right]}{1-\epsilon} + \frac{\val(D^S)}{\epsilon}$$
$$= \left(\frac{1}{1-\epsilon}+\frac{1}{\epsilon}\right) \val(D^S) + \frac{1}{1-\epsilon}\E_{v^S \leftarrow D^S}\left[\rev(D^X|v^S)\right],$$
as desired.
\end{proof}

 We can now provide our approximate core
decomposition by combining sub domain stitching (Lemma~\ref{lem:sds}) and
approximate marginal mechanism (Lemma \ref{lem:(Approximate-Marginal-Mechanism)}).

\begin{lemma}\label{lem:(Approximate-Core-Decomposition)}(``Approximate Core
Decomposition'') For any distribution $D$ that is subadditive over independent items,
and any $0<\epsilon<1$, 

\[
\textsc{Rev}\left(D\right)\leq\left(\frac{1}{\epsilon}+\frac{1}{1-\epsilon}\right)\textsc{Val}\left(D_{\emptyset}^{C}\right)+\frac{1}{1-\epsilon}\sum_{A\subseteq\left[n\right]}p_{A}\textsc{Rev}\left(D_{A}^{T}\right).
\]

\end{lemma}

In particular, for $\epsilon=1/2$, we have
\[
\textsc{Rev}\left(D\right)\leq4\textsc{Val}\left(D_{\emptyset}^{C}\right)+2\sum_{A\subseteq\left[n\right]}p_{A}\textsc{Rev}\left(D_{A}^{T}\right)
\]

\begin{proof}
By the Approximate Marginal Mechanism Lemma (Lemma \ref{lem:(Approximate-Marginal-Mechanism)}),
\begin{eqnarray*}
\textsc{Rev}\left(D_{A}\right) & \leq & \left(\frac{1}{\epsilon}+\frac{1}{1-\epsilon}\right)\textsc{Val}\left(D_{A}^{C}\right)+\frac{1}{1-\epsilon}\textsc{Rev}\left(D_{A}^{T}\right)
\end{eqnarray*}

Also, for any $A\subseteq\left[n\right]$, 
\[
\textsc{Val}\left(D_{A}^{C}\right)\leq\textsc{Val}\left(D_{\emptyset}^{C}\right)
\]
Finally, by sub-domain stitching (Lemma \ref{lem:sds}): 
\begin{eqnarray*}
\textsc{Rev}\left(D\right) & \leq & \sum_{A\subseteq\left[n\right]}p_{A}\textsc{Rev}\left(D_{A}\right)\\
 & \leq & \sum_{A\subseteq\left[n\right]}p_{A}\left(\left(\frac{1}{\epsilon}+\frac{1}{1-\epsilon}\right)\textsc{Val}\left(D_{A}^{C}\right)+\frac{1}{1-\epsilon}\textsc{Rev}\left(D_{A}^{T}\right)\right)\\
 & \leq & \left(\frac{1}{\epsilon}+\frac{1}{1-\epsilon}\right)\textsc{Val}\left(D_{\emptyset}^{C}\right)+\frac{1}{1-\epsilon}\sum_{A\subseteq\left[n\right]}p_{A}\textsc{Rev}\left(D_{A}^{T}\right)
\end{eqnarray*}
\end{proof}

\subsection{\label{sub:Core}Core}

Here, we show how to bound $\textsc{Val}(D_{\emptyset}^{C})$ using
$\max\{\srevstar_{\vec{q}}(D),\textsc{BRev}(D)\}$. We use a concentration
result due to Schechtman~\cite{Sch99-concentration_results} that
first requires a definition. Intuitively, the definition below says that a distribution is $c$-Lipschitz if changing the ``private information'' for a single item cannot change the buyer's value for any set by more than $c$. Moreover, adding/removing a single item cannot change the buyer's value for any set by more than $c$.
\begin{defn}
Let $D^{\vec{x}}$ denote a distribution of private information, $V$
denote a valuation function $V(\vec{x},\cdot)$, and $D$ denote the
distribution that samples $\vec{x}\leftarrow D^{\vec{x}}$ and outputs
the function $v(\cdot)=V(\vec{x},\cdot)$. Then $D$ is $c-${\em Lipschitz}
if for all $\vec{x},\vec{y}$, and sets $S$ and $T$ we have:
\[
\left|V\left(\vec{x},S\right)-V\left(\vec{y},T\right)\right|\leq c\cdot\left(\left|S\cup T\right|-|S\cap T|+\left|\left\{ i\in S\cap T\colon x_{i}\neq y_{i}\right\} \right|\right).
\]

\end{defn}
Before applying Schechtman's theorem, we show that $D_{\emptyset}^{C}$
is $t$-Lipschitz (recall that $t$ is the cutoff between core and
tail).

\begin{lemma}\label{lem:lipschitz}Let $D$ be any distribution that is subadditive
over independent items where each $v(\{i\})\in[0,c]$ with probability
$1$. Then $D$ is $c$-Lipschitz.

\end{lemma}
\begin{proof}
For any $\vec{x},\vec{y},S,T$, let $U=\{i\in S\cap T|x_{i}=y_{i}\}$.
Because of no externalities, we must have $V(\vec{x},U)=V(\vec{y},U)$,
which we will denote by $B$. By monotonicity, we must have $V(\vec{x},S),V(\vec{y},T)\geq B$.

Now, by subadditivity and the fact that each $V(\vec{x},\{i\})\leq c$,
we have $V(\vec{x},S)\leq c(|S|-|U|)+B$. Similarly, we have $V(\vec{y},T)\leq c(|T|-|U|)+B$.

It's also clear that $|S|-|U|\leq|S\cup T|-|S\cap T|+|\{i\in S\cap T:x_{i}\neq y_{i}\}|$ (the RHS is just rewriting the size of $S\cup T - U$),
and that $|T|-|U|\leq|S\cup T|-|S\cap T|+|\{i\in S\cap T:x_{i}\neq y_{i}\}|$.

So by everything above, we must have $V(\vec{x},S),V(\vec{y},T)\leq B+c(|S\cup T|-|S\cap T|+|\{i\in S\cap T:x_{i}\neq y_{i}\}|)$.
Therefore $V(\vec{x},S),V(\vec{y},S)\in[B,B+c(|S\cup T|-|S\cap T|+|\{i\in S\cap T:x_{i}\neq y_{i}\}|)]$,
completing the proof.
\end{proof}

\begin{corollary}
\label{cor:Lipschitz}$D_{\emptyset}^{C}$ is $t$-Lipschitz.
\end{corollary}
Now we state Schechtman's theorem and apply it to bound $\textsc{Val}(D_{\emptyset}^{C})$. 
\begin{theorem}
\label{thm:subadditive-concentration}(\cite{Sch99-concentration_results})
Suppose that $D$ is a distribution that is subadditive over independent
items and $c$-Lipschitz. Then for any parameters $q,a,k>0$, 
\[
\Pr_{v\leftarrow D}\left[v\left(\left[n\right]\right)\geq\left(q+1\right)a+k\cdot c\right]\leq\Pr\left[v\left(\left[n\right]\right)\leq a\right]^{-q}q^{-k}
\]

In particular, if $a$ is the median of $v\left([n]\right)\mid_{v\leftarrow D}$
and $q=2$, we have 
\[
\Pr_{v\leftarrow D}\left[v\left(\left[n\right]\right)\geq3a+k\cdot c\right]\leq4\cdot2^{-k}
\]
\end{theorem}
\begin{corollary}
\label{cor:concentration}Suppose that $D$ is a distribution that
is subadditive over independent items and $c$-Lipschitz. If $a$
is the median of $v\left(\left[n\right]\right)\mid_{v\leftarrow D}$,
then $\E_{v\leftarrow D}\left[v\left(\left[n\right]\right)\right]\leq3a+4c/\ln2$.\end{corollary}
\begin{proof}
$\E\left[v\left(\left[n\right]\right)\right]=\int_{0}^{\infty}\Pr\left[v\left(\left[n\right]\right)>y\right]dy$.
We can upper bound this using the minimum of $1$ and the bound provided
in Theorem \ref{thm:subadditive-concentration} to yield:

\[
\E\left[v\left(\left[n\right]\right)\right]\leq3a+\int_{0}^{\infty}4\cdot2^{-y/c}dy=3a+4c/\ln2
\]
\end{proof}
\begin{proposition}
\label{prop:core}$\textsc{Val}(D_{\emptyset}^{C})\leq6\textsc{BRev}+4t/\ln2$.\end{proposition}
\begin{proof}
Since $a$ is the median of $v\left(\left[n\right]\right)$, we can
set price $a$ on the grand bundle and extract revenue at least $a/2$.
Therefore, $\textsc{BRev}\geq a/2$. By Corollary~\ref{cor:concentration} and~\ref{cor:Lipschitz}, we have $\val(D^C_\emptyset) \leq 3a + 4t/\ln(2)$, so $\val(D^C_\emptyset) \leq 6\textsc{BRev} + 4t/\ln(2)$. 
\end{proof}
Finally, if the cutoff $t$ is not too large, we can recover a constant
fraction of it by selling each item separately.
\begin{lemma}
\label{lem:t}
$\srevstar_{\vec{p}}\geq t\cdot p_{\emptyset}\left(1-p_{\emptyset}\right)$.
In particular, if we choose $t$ so that $p_{\emptyset}=1/2$, then
$\srevstar_{\vec{p}}\geq t/4$.
\end{lemma}

\begin{proof}
Clearly $\textsc{Rev}_{p_{i}}(D_{i})\geq p_{i}t$, as we could set
a price of $t$ for item $i$. So $\srevstar_{\vec{p}}=p_{\emptyset}\sum_{i}\rev_{p_{i}}(D_{i})\geq p_{\emptyset}t\sum_{i}p_{i}$
Finally, we observe that $\sum_{i}p_{i}$ is exactly the expected
number of items in the tail, and $p_{\emptyset}$ is the probability
that zero items are in the tail. So we clearly have $\sum_{i}p_{i}\geq1-p_{\emptyset}$.
\end{proof}

Combining Proposition~\ref{prop:core} and Lemma~\ref{lem:t} then yields:
\begin{proposition}\label{thm:core}\[
\textsc{Val}(D_{\emptyset}^{C})\leq6\textsc{BRev}+24\srevstar_{\vec{p}}\mbox{.}
\]
\end{proposition}
\begin{proof}
By Proposition~\ref{prop:core}, $\val(D^C_\emptyset) \leq 6\textsc{BRev} + 4t/\ln(2)$. By Lemma~\ref{lem:t}, $t \leq 4 \srevstar_{\vec{p}}$. As $16/\ln(2) \leq 24$, we get the proposition as desired.
\end{proof}

\subsection{\label{sub:Tail}Tail}

We now show that the revenue from the tail can be approximated by
$\srevstar_{\vec{q}}$. We begin by proving a much weaker bound on
the optimum revenue for any distribution of subadditive valuations
over independent items:

\begin{lemma}

\label{lem:weak_bound_SRev}
\[
\textsc{Rev}\left(D\right)\leq6n^{\log_{2}6}\sum_{i}\textsc{Rev}\left(D_{i}\right)\mbox{.}
\]

\end{lemma}
\begin{proof}
Babaioff et al. \cite{BabaioffILW14} prove that $\textsc{Rev}\leq n\sum_{i}\textsc{Rev}\left(D_{i}\right)$
for an additive buyer by recursively reducing the number of items
by one at each step. Unfortunately, each step of the induction uses
the Marginal Mechanism Lemma; when applying the approximate variant
for subadditive valuations, we would incur an exponential factor. 

Instead, we use a slightly more complicated argument along the lines
of Hart and Nisan \cite{HartN12} that halves the number of items
in each step. Let $S$ and $X$ be a partition of $\left[n\right]$
into subsets of size at most $\lceil n/2\rceil$. Let $D_{S\geq X}$
be the distribution over valuations which is the same as $D$ whenever
$v\left(S\right)\geq v\left(T\right)$, and has valuation zero otherwise.
Similarly, let $D_{S<X}$ be the distribution which is equal to $D$
on $v\left(S\right)<\left(T\right)$. Then by sub-domain stitching
(Lemma \ref{lem:sds}) we have,
\begin{equation}
\textsc{Rev}\left(D\right)\leq\textsc{Rev}\left(D_{S\geq X}\right)+\textsc{Rev}\left(D_{S<X}\right)\mbox{.}\label{eq:split_Dst_Dts}
\end{equation}
Now, by the Approximate Marginal Mechanism Lemma, 
\begin{equation}
\textsc{Rev}\left(D_{S\geq X}\right)\leq\left(\frac{1}{\epsilon}+\frac{1}{1-\epsilon}\right)\textsc{Val}\left(D^{X}_{S\geq X}\right)+\frac{1}{1-\epsilon}\E_{v^{X}\leftarrow D^{X}_{S\geq X}}\left[\textsc{Rev}\left(D^{S}_{S\geq X}\mid v^{X}\right)\right]\label{eq:splitS>T}
\end{equation}
One mechanism for selling items in $S$ is to sample $v^{X}\leftarrow D^{X}_{S\geq X}$,
and then use a mechanism that achieves $\textsc{Rev}\left(D^{S}_{S\geq X}\mid v^{X}\right)$.
Thus we have, 
\begin{gather}
\textsc{Rev}\left(D^{S}\right)\geq\E_{{v}^{X}\leftarrow D^{X}_{S\geq X}}\left[\textsc{Rev}\left(D^{S}_{S\geq X}\mid v^{X}\right)\right]\mbox{.}\label{eq:RevDs_RevDs_given_T}
\end{gather}
Another way to sell items in $S$ is to again sample $v^{X}\leftarrow D^{X}_{S\geq X}$,
and offer the entire bundle for price $v^{X}\left(X\right)$. Therefore
we also have,
\begin{gather}
\textsc{Rev}\left(D^{S}\right)\geq\E_{v\leftarrow D}\left[v\left(X\right)\mid\Big(v\left(S\right)\geq v\left(X\right)\Big)\right]=\E_{v\leftarrow D_{S\geq X}}\left[v\left(X\right)\right]=\textsc{Val}\left(D^{X}_{S\geq X}\right)\mbox{.}\label{eq:RevDs_ValDt}
\end{gather}
Combining equations (\ref{eq:splitS>T})-(\ref{eq:RevDs_ValDt}),
we have 
\[
\textsc{Rev}\left(D_{S\geq X}\right)\leq\left(\frac{1}{\epsilon}+\frac{2}{1-\epsilon}\right)\textsc{Rev}\left(D^{S}\right)
\]
By symmetry, the same holds for $\textsc{Rev}\left(D_{S<X}\right)$
and $\textsc{Rev}\left(D^{X}\right)$. Therefore using (\ref{eq:split_Dst_Dts}),
\[
\textsc{Rev}\left(D\right)\leq\left(\frac{1}{\epsilon}+\frac{2}{1-\epsilon}\right)\left(\textsc{Rev}\left(D^{S}\right)+\textsc{Rev}\left(D^{X}\right)\right)
\]
Applying the recursion $\lceil\log n\rceil$ times, we have 
\[
\textsc{Rev}\left(D\right)\leq\left(\frac{1}{\epsilon}+\frac{2}{1-\epsilon}\right)^{\log_{2}n+1}\sum_{i}\textsc{Rev}\left(D_{i}\right)=\left(\frac{1}{\epsilon}+\frac{2}{1-\epsilon}\right)n^{\log_{2}\left(\frac{1}{\epsilon}+\frac{2}{1-\epsilon}\right)}\sum_{i}\textsc{Rev}\left(D_{i}\right)
\]
Choosing $\epsilon=1/2$ yields $\left(\frac{1}{\epsilon}+\frac{2}{1-\epsilon}\right)=6$.
\end{proof}

Note that in Lemma \ref{lem:weak_bound_SRev}, $\sum_{i}\textsc{Rev}(D_{i})$
is \emph{not} the same as $\textsc{SRev}(D)$ as the buyer is not
necessarily additive. In fact, they can be off by a factor of $n$
(in the case of a unit-demand buyer). Nonetheless, this weak bound
suffices for our analysis of the tail, which is summarized in Proposition~\ref{prop:tail}
below. Essentially, the proposition amplifies the bound in Lemma~\ref{lem:weak_bound_SRev}
greatly by making use of the fact that it is unlikely to see multiple
items in the tail. 
\begin{proposition}\label{prop:tail}
Recall that $p_{i}=\Pr[v(\{i\})>t]$, and $p_{\emptyset}=\prod_{i}(1-p_{i})$.
Then 
\[
\sum_{A}p_{A}\rev(D_{A}^{T})\leq\frac{6}{p_{\emptyset}}\left(1+7\ln(1/p_{\emptyset})+6\ln(1/p_{\emptyset})^{2}+\ln(1/p_{\emptyset})^{3}\right)\cdot\srevstar_{\vec{p}}
\]

In particular, if we choose $t$ so that $p_{\emptyset}=1/2$, then
$\sum_{A}p_{A}\rev(D_{A}^{T})\leq109\cdot\srevstar_{\vec{p}}$
\end{proposition}
\begin{proof}
Our proof builds on the intuition that the number of items in the
tail is typically very small. By Lemma \ref{lem:weak_bound_SRev},
we have that 
\begin{eqnarray}
\sum_{A\subseteq\left[n\right]}p_{A}\textsc{Rev}\left(D_{A}^{T}\right) & \leq & \sum_{A\subseteq\left[n\right]}p_{A}6\left|A\right|^{\log_{2}6}\sum_{i\in A}\textsc{Rev}\left(D_{i}^{T}\right)\nonumber \\
 & = & 6\sum_{i\in[n]}p_{i}\textsc{Rev}\left(D_{i}^{T}\right)\sum_{A\ni i}\left|A\right|^{\log_{2}6}p_{A}/p_{i}\label{eq:ub_pr_RevA}
\end{eqnarray}
For any $i$, the expression $\sum_{A\ni i}\left|A\right|p_{A}/p_{i}$
is also the expected number of items in the tail, conditioning on
$i$ being in the tail. Similarly, $\sum_{A\ni i}\left|A\right|^{\log_{2}6}p_{A}/p_{i}$
is the expected value of $\text{(\# items)}^{\log_{2}6}$. Let $b_{j}$
be the indicator random variable that is $1$ whenever item $j$ is
in the tail. Noting that $\log_{2}6<3$ and each $b_{j}$ is $1$
with probability exactly $p_{j}$ and the $b_{j}$'s are independent,
we have:
\begin{eqnarray}
\sum_{A\ni i}\left|A\right|^{\log_{2}6}p_{A}/p_{i} & \leq & \E_{b_{j}}\left[\left(1+\sum_{j\neq i}b_{j}\right)^{3}\right]\nonumber \\
 & = & \E_{b_{j}}\left[1+3\left(\sum_{j\neq i}b_{j}\right)+3\left(\sum_{j\neq i}b_{j}\right)^{2}+\left(\sum_{j\neq i}b_{j}\right)^{3}\right]\nonumber \\
 & = & 1+3\E\left[\sum_{j\neq i}b_{j}\right]+3\E\left[\sum_{j\neq i}b_{j}^{2}+\sum_{k\neq j\neq i}b_{j}b_{k}\right]\nonumber \\
 &  & +\E\left[\sum_{j\neq i}b_{j}^{3}+3\sum_{k\neq j\neq i}b_{j}^{2}b_{k}+\sum_{l\neq k\neq j\neq i}b_{j}b_{k}b_{l}\right]\nonumber \\
 & = & 1+7\E\left[\sum_{j\neq i}b_{j}\right]+6\E\left[\sum_{k\neq j\neq i}b_{j}b_{k}\right]+\E\left[\sum_{l\neq k\neq j\neq i}b_{j}b_{k}b_{l}\right]\label{eq:bj_in_01}\\
 & \leq & 1+7\sum_{j\neq i}p_{j}+6\left(\sum_{j\neq i}p_{j}\right)^{2}+\left(\sum_{j\neq i}p_{j}\right)^{3}\label{eq:leq_sum_pj}
\end{eqnarray}
(\ref{eq:bj_in_01}) follows because $b_{j}\in\left\{ 0,1\right\} $.
We continue to bound the last line as a function of just $p_{\emptyset}$.
Note that $e^{-\sum_{i}p_{i}}\geq\prod_{i}(1-p_{i})=p_{\emptyset}$,
and therefore we have $\sum_{i}p_{i}\leq\ln(1/p_{\emptyset})$. Combining
with (\ref{eq:ub_pr_RevA}) and (\ref{eq:leq_sum_pj}), we have:
\[
\sum_{A\subseteq\left[n\right]}p_{A}\textsc{Rev}\left(D_{A}^{T}\right)\leq6\left(1+7\ln(1/p_{\emptyset})+6\ln(1/p_{\emptyset})^{2}+\ln(1/p_{\emptyset})^{3}\right)\cdot\sum_{i\in\left[n\right]}p_{i}\textsc{Rev}\left(D_{i}^{T}\right)
\]

Now, we have to interpret $p_{i}\textsc{Rev}(D_{i}^{T})$. We claim
that in fact this is exactly $\textsc{Rev}_{p_{i}}(D_{i})$. Why?
It's clear that the optimal reserve for $D_{i}^{T}$ is at least $t$,
as $D_{i}^{T}$ is not supported below $t$. It's also easy to see
that for any reserve $r_{i}\geq t$, that the revenue obtained by
selling to $D_{i}^{T}$ is exactly $r_i \cdot \Pr[v(\{i\})>r_{i}]/p_{i}$, and
therefore the same $r_{i}\geq t$ that is optimal for $D_{i}$ is
also optimal for $D_{i}^{T}$, and $p_{i}\textsc{Rev}(D_{i}^{T})=\textsc{Rev}_{p_{i}}(D_{i})$.
Therefore,
\[
\sum_{A\subseteq\left[n\right]}p_{A}\textsc{Rev}\left(D_{A}^{T}\right)\leq6\left(1+7\ln(1/p_{\emptyset})+6\ln(1/p_{\emptyset})^{2}+\ln(1/p_{\emptyset})^{3}\right)\cdot\sum_{i\in\left[n\right]}\textsc{Rev}_{p_{i}}\left(D_{i}\right)
\]
Plug in $\srevstar_{\vec{p}}=p_{\emptyset}\sum_{i\in\left[n\right]}\textsc{Rev}_{p_{i}}\left(D_{i}\right)$
to complete the proof.
\end{proof}

Note that Theorem~\ref{thm:main} is now a corollary of Proposition~\ref{prop:tail},
Proposition~\ref{thm:core}, and Lemma \ref{lem:(Approximate-Core-Decomposition)}
(setting $\epsilon=1/2$). That the desired $\vec{q}$ can be computed efficiently is easy to see: simply do a binary search over cutoffs $t$ until we find one that induces $p_\emptyset = 1/2$. It is also easy to find an item pricing that guarantees revenue at least $\srev^*_{\vec{q}}$: for each item $i$, simply find the optimal reserve for $D_i$, subject to that reserve being at least $t$. Finally, notice that the only bundle price we ever need to set to obtain our guarantee is the median of $v([n])$, when $v(\cdot) \leftarrow D^C_{\emptyset}$. It is also easy to see that our bounds degrade smoothly if we set a price that only approximates the median instead. For a discussion of exactly what access to $D$ suffices in order for these prices/cutoffs to be truly easy to find, we refer the reader to~\cite{BabaioffILW14}. We note here just that it should be clear that any reasonable, even minimal, access to $D$ does indeed suffice.


\section{Simple Auctions and Approximate Revenue
Monotonicity}
\label{sec:monotonicity}

In this section we explore the rich connection between approximately
optimal simple auctions, and approximate revenue monotonicity. By
approximate revenue monotonicity, we formally mean the following:
\begin{defn}
We say that a class of distributions is {\em $\alpha$-approximately
revenue monotone} if for any two distributions $D$ and $D^{+}$
in that class such that $D^{+}$ first-order stochastically dominates
$D$ (recall that we $D^+$ first-order stochastically dominates $D$ if they can be coupled so that when we sample $v^+$ from $D^+$ and $v$ from $D$ we have $v^+(S) \geq v(S)$ for all $S$),
$\alpha\cdot\textsc{Rev}\left(D^{+}\right)\geq\textsc{Rev}\left(D\right)$.
\end{defn}

In the rest of the section we observe that subadditive valuations
over independent items are $\alpha$-approximately monotone for some
constant factor (Subsection \ref{sub:Approximate-monotonocity-for-subadditive}).
We also note that a (significantly) tighter approximate monotonicity
would yield a better factor of approximation in Theorem~\ref{thm:main}
(Subsection \ref{sub:Better-approximate-monotonocity-implies}). Finally,
for the class of (possibly correlated) additive valuations over $n$
items, we prove a reduction from approximate revenue monotonicity
to approximately optimal simple auctions (that loses a factor of $n$).
Then we use an infinite gap between $\max\left\{ \textsc{BRev},\textsc{SRev}\right\} $
and $\textsc{Rev}$ for two correlated items due to Hart and Nisan \cite{HartN13}
to prove an infinite lower bound on approximate revenue monotonicity
(Subsection \ref{sub:Correlated-distributions-are-nonmonotone}).

\subsection{\label{sub:Approximate-monotonocity-for-subadditive}Approximately
optimal simple auctions imply approximate monotonicity}

As a corollary of our main theorem (Theorem \ref{thm:main}) we deduce
constant-factor approximate monotonicity for subadditive valuations
over independent items:
\begin{corollary}
The class of subadditive valuations over independent items is $338$-approximately
monotone.
\end{corollary}
Similarly, from the $6$-approximation of Babaioff et al. for additive
yields.
\begin{corollary}
The class of additive valuations over independent items is $6$-approximately
monotone.\end{corollary}
\begin{proof}
For additive functions, $\textsc{BRev}$ and $\textsc{SRev}$ constitute
of separate Myerson's auctions, and are therefore revenue monotone.
Thus,
\begin{flalign*}
6\textsc{Rev}\left(D^{+}\right) & \geq6\max\left\{ \textsc{BRev}\left(D^{+}\right),\textsc{SRev}\left(D^{+}\right)\right\} \\
 & \geq6\max\left\{ \textsc{BRev}\left(D\right),\textsc{SRev}\left(D\right)\right\} \geq\textsc{Rev}\left(D\right)
\end{flalign*}

For subadditive functions, $\textsc{SRev}(D)$ is no longer monotone,
but $\srevstar_{\vec{q}}(D)$ is. This is because $\srev_{q}(D_{i})$
is clearly monotone, and $\srevstar_{\vec{q}}$ is just a (scaled)
sum of $\srev_{q_{i}}(D_{i})$. So we get that there exists a $\vec{q}$
such that:

\begin{flalign*}
338\textsc{Rev}\left(D^{+}\right) & \geq338\max\left\{ \textsc{BRev}\left(D^{+}\right),\srevstar_{\vec{q}}\left(D^{+}\right)\right\} \\
 & \geq338\max\left\{ \textsc{BRev}\left(D\right),\srevstar_{\vec{q}}\left(D\right)\right\} \geq\textsc{Rev}\left(D\right)
\end{flalign*}
\end{proof}

\subsection{\label{sub:Better-approximate-monotonocity-implies}Approximate monotonicity
implies approximately optimal simple auctions}

A closer look at the proof of our main theorem also yields the converse
of the above corollaries, namely: a tighter approximate monotonicity
for subadditive valuations would yield an improved factor of approximation
by simple auctions, as well as a simpler proof. 
\begin{corollary}
If the class of subadditive valuations over independent items is $\alpha$-approximately
monotone, then 
\[
\textsc{Rev}\leq\alpha\Big(\left(37\alpha+24\right)\textsc{SRev}+6\textsc{BRev}\Big)
\]
\end{corollary}
\begin{proof}
(Sketch)

Recall that in the proof of the Approximate Marginal Mechanism Lemma
(Lemma \ref{lem:(Approximate-Marginal-Mechanism)}), we made use of
Lemma~\ref{lem:delta} to bound the gap between $\rev(D^{+})$ and
$\rev(D)$, where $D^{+}$ first-order stochastically dominated $D$.
Instead of the $\epsilon$-truthful to truthful reduction, we could
derive $\alpha\rev(D^{+})\geq\textsc{Rev}\left(D\right)$ from approximate
monotonicity. Then, we can directly apply Lemma \ref{lem:HNmarginal}
to get: 
\[
\textsc{Rev}\left(D\right)\leq\alpha\left(\textsc{Val}\left(D^{S}\right)+\E_{v^{S}\sim D^{S}}\left[\textsc{Rev}\left(D^{X}\mid\mathbf{v}^{S}\right)\right]\right)
\]
Instead of 
\[
\textsc{Rev}\left(D\right)\leq4\textsc{Val}\left(D^{S}\right)+2\E_{v^{S}\sim D^{S}}\left[\textsc{Rev}\left(D^{X}\mid\mathbf{v}^{S}\right)\right]
\]
If $\alpha\leq2$, this indeed yields a tighter approximation. 
\end{proof}

\subsection{\label{sub:Correlated-distributions-are-nonmonotone}Correlated Distributions
are not Approximately Monotone}

So far we've shown that (for some valuation classes) approximately
optimal simple mechanisms imply approximate revenue-monotonicity.
Are all classes approximately revenue-monotone? In this subsection
we provide a reduction from an instance where $\max\left\{ \srev,\brev\right\} $
does not approximate $\rev$ to show an infinite non-monotonicity
for correlated items. We first prove a reduction from gaps between
$\brev$ and $\rev$ to non-monotonicity.
\begin{proposition}
\label{prop:monotone}Let $D$ be a distribution over subadditive
valuations for $n$ items for which $\rev(D)>c\cdot\brev(D)$. Then
any class of distributions containing $D$ and all single-dimensional
distributions%
\footnote{A distribution is single-dimensional if for all $v$ in its support,
$v(S)=c|S|$ for some value $c$.%
} is not $(c/n)$-approximately revenue monotone.\end{proposition}
\begin{proof}
We define $D^{+}$ as follows. First, sample $v\leftarrow D$. Then
let $i^{*}=\arg\max_{i}\{v(\{i\})\}.$ Now, set $v^{+}(S)=v(\{i^{*}\})\cdot|S|$
for all $|S|$. By subadditivity, it's clear that $D^{+}$ first order
stochastically dominates $D$. Now, however, $D^{+}$ is a single-dimensional
distribution, meaning that $\brev(D^{+})=\rev(D^{+})$~\cite{Myerson81, RileyZ83}.
Finally, we just need to compare $\brev(D)$ to $\brev(D^{+})$. 

Note that by monotonicity, we have $v^{+}([n])\leq n\cdot v([n])$
for all $v,v^{+}$. Therefore, for any price $p$, if $v^{+}([n])>p,$
$v([n])>p/n$. This immediately implies that $\brev(D)\geq\brev(D^{+})/n$:
let $p$ be the optimal reserve for the grand bundle under $D^{+}$,
then setting price $p/n$ sells with at least the same probability
under $D$. Putting both observations together, we see that: $\rev(D)>c\brev(D)\geq(c/n)\brev(D^{+})=(c/n)\rev(D^{+})$,
meaning that any class containing $D$ and $D^{+}$ is not $(c/n)$-approximately
monotone.
\end{proof}
We apply Proposition~\ref{prop:monotone} to a theorem of Hart and
Nisan.
\begin{theorem}
\label{thm:infy-gap-brev_rev}(Hart and Nisan \cite{HartN13}) There
exists a distribution $D$ over correlated additive valuations for
two items such that $\textsc{BRev}\left(D\right)\leq1/2$, and $\textsc{Rev}\left(D\right)=\infty$.\end{theorem}
\begin{corollary}
There exist distributions $D$ and $D^{+}$ over over correlated additive
valuations for two items such that $D^{+}$ first-order stochastically
dominates $D$, $\textsc{Rev}\left(D^{+}\right)=1$, and yet $\textsc{Rev}\left(D\right)=\infty$.
Therefore, the class of correlated additive valuations for two items
is not $c$-approximately revenue monotone for any finite $c$.
\end{corollary}

\section{Approximate Marginal Mechanism and Core Decomposition
for Many Bidders}
\label{app:many}

In this section we provide a complete proof of the Marginal Mechanism
and Core Decomposition lemmas that apply to many bidders. The statements
are essentially the same, but require more extensive notation which
we develop in the next subsection. 

The main technical challenge is extending Lemma \ref{lem:delta} to
many buyers; recall that this is the lemma that lower bounds the revenue
from a distribution $D^{+}$ of additive-across-a-partition valuations,
in terms of the revenue from the original distribution $D$ and the
expected difference $\overline{\delta}$ between $D$ and $D^{+}$.
I.e. given a black-box mechanism for $D$, we would like to create
a new mechanism for $D^{+}$. How can we preserve incentive compatibility
when the buyers have different incentives? In Section \ref{sub:cdl}
we do this for a single buyer by giving the buyer the outcome (allocation
and price), of the possible outcomes for all types in the original
mechanism, which is optimal for her new valuation. Incentive compatibility
is now guaranteed, and we simply need to bound the revenue. 

When multiple buyers are involved there is a problem with this approach:
incentive compatibility of any buyer depends also on the distribution
of types reported by other buyers; thus we cannot simply let any buyer
report any type she wants. To overcome this challenge, we use a technique
due to \cite{BeiH11,HartlineKM11,DaskalakisW12} which guarantees
that each buyer observes the correct distribution of types on the
other buyers. For buyer $i$, we sample an additional $r-1$ {\em replica}
types from $D_{i}^{+}$, and $r$ {\em surrogate} types from $D_{i}$.
Given the real buyer's type, we create (in an incentive compatible
manner) a complete matching between replicas and surrogates. Since
the real buyer's type is sampled from the same distribution as the
replicas, she is equally likely to be matched to any of the surrogates.
Thus the distribution over the surrogate type that is matched to the
buyer $i$ is exactly the original distribution $D_{i}$; i.e. all
other buyers observe the ``correct'' distribution. The mechanism
is now Bayesian incentive compatible, and further analysis shows that
the lower bound on the revenue is preserved.

\subsection{Notation and statement}

There are $m$ bidders and $n$ items. We now say that an item $i$
is in the tail if there is \emph{any} buyer who values item $i$
above the core-tail cutoff, and item $i$ is in the core if every
buyer values it below the cutoff. $D$ is now the joint distribution
of all buyers valuation functions. We study Bayesian Incentive Compatible
(BIC) mechanisms (and $\eta$-BIC mechanisms). A mechanism is BIC if it is in every buyer's best interest to report truthfully, conditioned on other buyers reporting
truthfully as well (and $\eta$-BIC if this holds up to an additive $\eta$). Formally, let $\phi_{j}(v)$ denote the (random)
allocation awarded to bidder $j$ when reporting type $v$; we slightly
abuse notation by letting $v\left(\phi\right)$ denote the expected
utility, over any randomness in the mechanism and the other bidders
sampling their types, that a bidder with type $v$ gains from a random
allocation $\phi$; let $p_{j}(v)$ denote the expected price paid
by bidder $j$ when reporting type $v$ over the same randomness.
Finally, a mechanism is BIC iff for all $j,v,v'$, $v(\phi_{j}(v))-p_{j}(v)\geq v(\phi_{j}(v'))-p_{j}(v')$. Let $\eta(\cdot)$ denote the additive function that satisfies $\eta(S) = \eta\cdot |S|$ (that is, the function has value $\eta$ per item). A mechanism is $\eta$-BIC iff for all $j,v,v'$, $v(\phi_{j}(v))-p_{j}(v) + \eta(\phi_j(v)) \geq v(\phi_{j}(v'))-p_{j}(v')$.
We use the following notation.
\begin{itemize}
\item $D_{j}$: The distribution of $v_{j}(\cdot)$, the valuation function
for bidder $j$. We assume that $D$ is a product distribution. That
is, $D=\times_{j}D_{j}$.
\item $D_{A}$: the distribution $D$, conditioned on $A$ being exactly
the set of items in the tail.
\item $D_{A}^{T}$: the distribution $D_{A}$ restricted just to items in
the tail (i.e. $A$). 
\item $D_{A}^{C}$: the distribution $D_{A}$ restricted just to items in
the core (i.e. $\bar{A}$). 
\item $p_{i}$: the probability that element $i$ is in the tail. 
\item $p_{A}$: the probability that $A$ is exactly the set of items in
the tail (note that $p_{\left\{ i\right\} }\neq p_{i}$).
\item $\textsc{Rev}\left(D\right)$: The maximum revenue obtainable via
a BIC mechanism from buyers with valuation profile drawn from $D$.
\item $\textsc{Val}\left(D\right)$=$\E_{v\leftarrow D}\left[\max_{S_{1}\sqcup\ldots\sqcup S_{m}}\{\sum_{j}v_{j}(S_{j})\}\right]$:
the expected welfare of the VCG mechanism when buyers are drawn from
$D$.
\item $\textsc{Rev}^{M}(D)$: The revenue of a mechanism $M$ when buyers
drawn from $D$ play truthfully.
\end{itemize}
We are finally ready to state the approximate core decomposition lemma
for many buyers:
\begin{lemma}\label{lem:marginalMany}
For any distribution $D=\times_{j}D_{j}$ with each $D_{j}$ subadditive
over independent items, and any $0<\epsilon<1$, 

\[
\rev(D)\leq\left(\frac{1}{\epsilon}+\frac{1}{1-\epsilon}\right)\val(D_{\emptyset}^{C})+\frac{1}{1-\epsilon}\sum_{S\subseteq[n]}p_{A}\rev(D_{A}^{T})
\]
\end{lemma}
\begin{proof}
Follows from Theorem \ref{thm:delta-many} below, together with the
arguments used in Section~\ref{sub:cdl} and in~\cite{BabaioffILW14}.
\end{proof}

\subsection{The mechanism}

First, we describe the reduction we will use (which is essentially
the same as the $\epsilon$-BIC to BIC reduction used in~\cite{DaskalakisW12},
but without some technical hardships since we aren't concerned with
runtime - our proof never actually runs this procedure). Below, $D^{+}$
denotes any product distribution that first-order stochastically dominates
$D$, and $\delta_{j}(\cdot)$ denotes the random function $v_{j}^{+}(\cdot)-v_{j}(\cdot)$
when couples $v^{+}$ and $v$ are sampled jointly from $D^{+}$ and
$D$. Note that $\delta_{j}(S)\geq0$ for all $\delta_{j},S$. We
will also abuse notation and refer to $\delta_{j}$ as the distribution
over $\delta_{j}(\cdot)$ as well (so we can write terms like $\val(\delta)$).

\paragraph{Phase 1, Surrogate Sale:}
\begin{enumerate}
\item {Let $M$ be any $\eta$-BIC mechanism for buyers from $D$. Multiply all
prices charged by $M$ by $(1-\epsilon)$ and call the new mechanism
$M^{\epsilon}$. Interpret the $\epsilon$ fraction of prices given
back as rebates.}
\item For each bidder $j$, create $r-1$ \emph{replicas} sampled i.i.d.
from $D_{j}^{+}$ and $r$ \emph{surrogates} sampled i.i.d. from
$D_{j}$. Let $r\rightarrow\infty.$
\item Ask each bidder to report $v_{j}(\cdot)$. 
\item Create a weighted bipartite graph with replicas (and bidder $j$)
on the left and surrogates on the right. The weight of an edge between
a replica (or bidder $j$) with type $r_{j}(\cdot)$ and surrogate
of type $s_{j}(\cdot)$, is the utility of $r_{j}$ for the expected
outcome of $M^{\epsilon}$ when reporting $s_{j}$. That is, the weight
of the edge is $r_{j}(\phi_{j}^{\epsilon}(s_{j}))-p_{j}^{\epsilon}(s_{j})$.
\item Compute a maximum matching and VCG prices in this bipartite graph;
we henceforth refer to it as the {\em VCG matching}. If a replica
(or bidder $j$) is unmatched in the VCG matching, add an edge to
a random unmatched surrogate. (Notice that some replicas may indeed
be unmatched if the gain negative utility from the allocation and
prices corresponding to some surrogates.) The surrogate selected for
bidder $j$ is whomever she is matched to.
\end{enumerate}

\paragraph*{Phase 2, Surrogate Competition:}
\begin{enumerate}
\item Let $s_{j}$ denote the surrogate chosen to represent bidder $j$
in phase one, and let $\vec{s}$ denote the entire surrogate profile
(i.e. the ones matched to the real buyers). Have the surrogates play
$M^{\epsilon}$. 
\item If bidder $j$ was matched to her surrogate through VCG, charge them
the VCG price and award them $M_{j}^{\epsilon}(\vec{s})$. Recall
that this has an allocation and price component; the price is added
onto the VCG price. If bidder $j$ was matched to a random surrogate
after VCG, award them nothing and charge them nothing.\end{enumerate}
\begin{theorem}
\label{thm:delta-many}Let $M'$ denote the mechanism of the process
above, starting from any $\eta$-BIC mechanism $M$ for consumers drawn from
$D$. Then $M'$ is BIC for consumers drawn from $D^{+}$. {Furthermore,
for any desired $\epsilon\in(0,1)$, we can have: $\rev^{M'}(D^{+})\geq(1-\epsilon)\cdot(\rev^{M}(D)-\val(\delta)/\epsilon-n\eta/\epsilon)$.}
\end{theorem}
In the theorem statement above, note that by $\val(\delta)$, we mean
the expected welfare of the VCG mechanism when buyers with types distributed
according to $\delta=\times_{j}\delta_{j}$ play, and recall that $n$ is the number of items.

\subsection{Proof outline}

The reduction is nearly identical to the reduction employed in~\cite{DaskalakisW12}
(which is itself inspired by~\cite{BeiH11, HartlineKM11}). We provide
complete proofs of all claims for completeness, noting that many of
these claims can also be found in~\cite{BeiH11,HartlineKM11,DaskalakisW12}.
We provide appropriate citations by each statement. Below is the proof
outline, taken from~\cite{DaskalakisW12}.
\begin{enumerate}
\item If bidder $j$ plays $M'$ truthfully, then the distribution of surrogates
matched to bidder $j$ is $D_{j}$. Therefore, the value of each edge
is calculated correctly as the expected utility of a replica with
type $r$ for being represented by a surrogate of type $s$ in $M^{\epsilon}$. 
\item Because each bidder is participating in a VCG auction for their surrogate,
and the value of each edge is calculated correctly, whenever all other
bidders tell the truth, it is in bidder $j$'s interest to tell the
truth as well. Therefore, $M'$ is BIC. 
\item The revenue made from bidder $j$ is at least the price paid by their
surrogate if bidder $j$ is matched in VCG, and 0 otherwise.
\item{ There exists a near-perfect matching that matches each replica to
a nearly-identical surrogate (modulo $\delta_{j}(\cdot)$). If VCG
used this matching, we would have $\rev^{M'}(D)=(1-\epsilon)\rev^{M}(D)$. }
\item {The rebates allow us to bound the revenue lost by selecting the VCG
matching instead of this near-perfect matching as a function of $\val(\delta), \eta,$
and $\epsilon$. }
\end{enumerate}
We proceed to state the formal claims associated with steps 1 through
5.

\subsubsection*{Step 1: The surrogate distributions}
\begin{lemma}
\label{lem:HKM1}(\cite{HartlineKM11}) If bidder $j$ plays $M'$
truthfully, then the distribution of the surrogate selected for bidder
$j$ is exactly $D_{j}$.\end{lemma}
\begin{proof}
Imagine sampling replicas and surrogates for bidder $j$ in the following
way instead. First, sample the $r$ types for the left-hand side of
the bipartite graph i.i.d. from $D_{j}^{+}$ and the $r$ types for
the right-hand side i.i.d. from $D_{j}$. Then, find the max-weight
matching between types, completing it by randomly adding edges from
unmatched nodes on the left to unmatched nodes on the right to form
a perfect matching. Then, declare all of the $r$ right-hand types
surrogates, randomly select one of the left-hand types to be bidder
$j$, and declare the remaining $r-1$ as replicas. Note that sampling
in this order yields the correct distribution of replicas, surrogates,
and bidder $j$, as long as bidder $j$ reports truthfully. Furthermore,
it is clear now that the distribution of the surrogate selected for
bidder $j$ after sampling in this order is exactly $D_{j}$: once
the matching is fixed, we simply pick a random left-hand type and
output its partner. So essentially we are drawing $r$ i.i.d. samples
from $D_{j}$ and selecting one at random. Clearly this is the same
distribution as $D_{j}$.
\end{proof}

\subsubsection*{Step 2: Bayesian incentive compatibility}
\begin{corollary}
(\cite{HartlineKM11}) $M'$ is BIC.\end{corollary}
\begin{proof}
Fix any bidder $j$ and assume all others report truthfully. By Lemma~\ref{lem:HKM1},
the distribution of all other surrogates matches $D_{-j}$ exactly,
so the weight of the edge between each replica (and bidder $j$) and
each surrogate correctly computes the value of that replica for being
represented by that surrogate in $M$. As bidder $j$ is just participating
in a truthful VCG mechanism against the replicas for the surrogates,
and all values are computed correctly (conditioned on other buyers
telling the truth), $M'$ is BIC. 
\end{proof}

\subsubsection*{Step 3: A good matching implies high revenue}
\begin{proposition}
\label{prop:3}(\cite{DaskalakisW12}) Conditioning on right-hand
types (surrogates) $\{s_{k}\}_{k\in[r]}$ and left-hand types (replicas
plus bidder $j$) $\{r_{k}\}_{k\in[r]}$, the expected payment of
bidder $j$ is at least 
\[
\sum_{k\mbox{\mbox{\,\ s.t.\,\ensuremath{s_{k}}\,\ is matched in VCG}}}p_{j}^{\epsilon}(s_{k})/r.
\]
\end{proposition}
\begin{proof}
Conditioned on the left and right-hand types being sampled, but having
not yet decided which left-hand type is bidder $j$, the surrogate
matched to bidder $j$ is just a random surrogate. So each surrogate
$s_{k}$ is matched to bidder $j$ with probability $1/r$. Furthermore,
bidder $j$ will pay the price $p_{j}^{\epsilon}(s_{k})$ whenever
his matched surrogate was selected by VCG (and not the random edges
afterwards). Therefore, bidder $j$ pays at least $\sum_{\{k|\mbox{\ensuremath{s_{k}\mbox{ is matched in VCG\}}}}}p_{j}^{\epsilon}(s_{k})/r$.
The reason this is not tight is because it does not count the additional
payments of the VCG mechanism.
\end{proof}

\subsubsection*{Step 4: Existence of a near-perfect matching}

Our next goal is to show that there exists a near-perfect matching
that only matches replicas and surrogates that are ``close.'' For
any $\gamma>0$ and two types $v$ and $v'$ drawn from $D_{j}$,
we say that $v$ and $v'$ are $\gamma$-equivalent if for all $S\subseteq[n]$,
there exists an integer $z(S)$ such that $\{v(S),v'(S)\}\subseteq[(1+\gamma)^{z(S)},(1+\gamma)^{z(S)+1})$.
It's easy to see that this defines an equivalence relation. For a
type $r_{j}$ drawn from $D_{j}^{+}$, let $r'_{j}$ denote it's couple
drawn from $D_{j}$. We say that two types drawn from $D_{j}^{+}$
are $\gamma$-equivalent if their couples are $\gamma$-equivalent,
and that $r_{j}$ drawn from $D_{j}^{+}$ is $\gamma$-equivalent
to $s_{j}$ drawn from $D_{j}$ if $r'_{j}$ and $s_{j}$ are $\gamma$-equivalent
(basically we are putting replicas in equivalence classes based on
their couples). 

The following lemma from \cite{HartlineKM11} bounds the number of
unmatched surrogates:
\begin{lemma}
\label{lem:HKMequiv} (Lemma 3.7 in~\cite{HartlineKM11}) The expected
number of unmatched surrogates in a maximal matching that only matches
equivalent replicas and surrogates when types are split into at most
$\beta$ possible equivalence classes is at most $O\left(\sqrt{\beta r}\right)$.
\end{lemma}
In the next lemma we use Lemma \ref{lem:HKMequiv} to lower-bound
the revenue obtained from the matched replicas.
\begin{lemma}
\label{lem:4}(Implicit in~\cite{HartlineKM11}) For any $\gamma>0$,
let $W$ denote any maximal matching of replicas to surrogates (for
all bidders) that only matches $\gamma$-equivalent types. Then as
$r\rightarrow\infty$, we get $\E[\sum_{j}\sum_{\{k|s_{jk}\text{ is matched in }W\}}p_{j}^{\epsilon}(s_{jk})]\geq(1-\gamma)\rev(M^{\epsilon})$. \end{lemma}
\begin{proof}
At a high level, the proof is straightforward: for a fixed equivalence
class, the distribution of the number of replicas and surrogates in
that class is the same. So as we take more and more i.i.d. samples,
the number of each concentrates very tightly around its expectation,
so not many types are unmatched. Showing this formally is somewhat
technical.

Let's focus on a specific equivalence class $C$ for a fixed bidder
$j$. There is some probability $q_{C}$ that a type drawn from $D_{j}$
lands in class $C$. Let $\rev^{q}(M^{\epsilon})$ denote the revenue
obtained by $M^{\epsilon}$ \emph{only counting payments by types in an equivalence class $C$ such that $q_C \geq q$ and $v([n])\leq 1/q$ for all $v\in C$.} 
It's clear that $\lim_{q\rightarrow0}\rev^{q}(M^{\epsilon})=\rev(M^{\epsilon})$,
as the revenue obtained from equivalence classes with $q_{C}=0$ is
exactly $0$ and the revenue obtained from all other equivalence classes
is eventually counted for sufficiently small $q$. So we can pick
a $q>0$ such that $\rev^{q}(M^{\epsilon})\geq(1-\gamma/2)\rev(M^{\epsilon})$.
Note now that there can only be finitely many equivalence classes
counted towards $\rev^{q}(M^{\epsilon})$ (in particular, at most
$1/q$ per bidder), and that the maximum payment by a type in any
such equivalence class is at most $1/q$ (by individual rationality). 

So now the probability that a surrogate or replica is sampled to be
in a counted equivalence class is at least $q$ (there must be at
least one such equivalence class) and there are at most $1/q$ equivalence
classes. So we may apply Lemma~\ref{lem:HKMequiv} to see that for
a single bidder, the expected number of unmatched surrogates from
equivalence classes that count is at most $O\left(\sqrt{r/q}\right)$.
As each such surrogate pays at most $1/q$, the total revenue lost
in expectation from unmatched surrogates in equivalence classes that
count is at most $O\left(1/\sqrt{rq^{3}}\right)$ (due to Proposition~\ref{prop:3}).
Summing up across all bidders, the revenue lost is at most $O\left(m/\sqrt{rq^{3}}\right)$.
In addition, the total revenue lost in expectation from unmatched
surrogates across all bidders from equivalence classes that didn't
count is clearly at most $(\gamma/2)\rev(M^{\epsilon})$ by choice
of $q$. So the total revenue lost from unmatched surrogates in this
matching is at most $(\gamma/2)\rev(M^{\epsilon})+O\left(m/\sqrt{rq^{3}}\right)$.
As $r\rightarrow\infty$, the second term approaches zero, completing
the proof.
\end{proof}

\subsubsection*{Step 5: The VCG matching is almost as good }

Combining Proposition~\ref{prop:3} and Lemma~\ref{lem:4} says the
following: if only this nearly-perfect matching was the one selected
by VCG, then we would know that $\rev^{M'}(D)$ was good. But for
all we know, VCG may choose to leave many surrogates unmatched if
it helps improve the welfare of the replicas. The key is to show that
not many surrogates can be unmatched, due to the rebates.

Let $V$ denote the VCG matching and $W$ denote the matching guaranteed
by Lemma~\ref{lem:4}. Then there is a disjoint set of augmenting
paths and cycles that transform $W$ into $V$. As $V$ is a max-weight
matching, \emph{all of these augmenting paths and cycles must have non-negative weight}. It
is easy to see that augmenting cycles do not change the set of matched
surrogates, and therefore do not affect the revenue. We therefore
want to study augmenting paths that unmatch a surrogate.

If an augmenting path unmatches $s_{jk}$, then no replica is receiving
the rebate awarded to $s_{jk}$ any more. Because VCG is choosing
the max-weight matching, it must be the case that the benefit of switching
every other edge along the path outweighs the cost of losing the rebate
awarded to $s_{jk}$. This is where we make use of the fact that each
replica is matched to a surrogate that is nearly identical to them,
except for an additive $\delta_{jk}(\cdot)$. Because $M$ is $\eta$-BIC,
we can bound the expected gain of switching a replica who is matched
to a nearly identical surrogate to any other surrogate using $\delta(\cdot)$ and $\eta(\cdot)$.
Therefore, each surrogate that gets unmatched by an augmenting path
``claims'' many replicas to be in its augmenting path. The argument
shows that in fact it takes several replicas to make a positive weight
augmenting path, and therefore not many surrogates can be unmatched. 
\begin{lemma}
(Ideas from~\cite{DaskalakisW12}) If $U_{j}$ denotes the set of
surrogates that are matched in $W$ but not $V$, and $T_{j}$ denotes
the set of surrogates matched in $V$ but not $W$, (for bidder $j$),
then $\E[\frac{1}{r}\sum_{j}\sum_{s\in U_{j}}p_{j}(s)-\sum_{s\in T_{j}}p_{j}(s)]\leq\val(\delta)/\epsilon +m\eta/\epsilon$. \end{lemma}
\begin{proof}
Consider any augmenting path that unmatches a surrogate $s$ in $W$
and (possibly) matches a new surrogate $s'$. For simplicity of notation,
if no new surrogate is matched, we let $s'$ denote a null type that
receives no items and pays no price to $M$. We break the contribution
of edges in this path into two parts, the first coming just from the
rebates awarded to the surrogates and the second coming from the allocation
and original price paid. It is easy to see that the contribution of
rebates to the weight of the augmenting path is exactly $\epsilon p_{j}(s)-\epsilon p_{j}(s')$.
Now we analyze the weight of the path coming from the second part.
We can compute the weight by summing over all replicas $r_{j}$ in
the path of their utility for their new surrogate minus their utility
for their old surrogate. Note that any augmenting path that unmatches
a surrogate cannot possibly add an edge to a replica who was unmatched
in $W$. Since $M$ is $\eta$-BIC, for any replica $r_{j}$ that was matched
to $s_{j}$ in $W$ and moved to $s_{j}'$ in $V$, we have:

\[
s_{j}(\phi_{j}(s_{j}))-p_{j}(s_{j}) + \eta(\phi_j(s_j))\geq s_{j}(\phi_{j}(s_{j}'))-p_{j}(s_{j})
\]

Using the fact that $r_{j}$ and $s_{j}$ are $\gamma$-equivalent (note that $\gamma$-equivalence implies that $(1+\gamma)r_j(S) \geq s_j(S) \geq (1-\gamma)r_j(S) - \delta_j(S)$ for all $S$), we get:

\[
(1+\gamma)r_{j}(\phi_{j}(s_{j}))-p_{j}(s_{j}) + \eta(\phi_j(s_j))\geq(1-\gamma)r_{j}(\phi_{j}(s_{j}'))-\delta_{j}(\phi_{j}(s_{j}'))-p_{j}(s_{j})
\]

And rearranging terms yields:

\[
r_{j}(\phi_{j}(s_{j}'))-p_{j}(s_{j})-r_{j}(\phi_{j}(s_{j}))+p_{j}(s_{j})\leq\gamma r_{j}(\phi_{j}(s_{j}))+\gamma r_{j}(\phi_{j}(s_{j}'))+\delta_{j}(\phi_{j}(s_{j}'))+\eta(\phi_j(s_j))
\]

Note that the left-hand side is exactly the increase in utility as
we move $r_{j}$ from $s_{j}$ to $s_{j}'$. So now we know that the
total increase in utility from moving all replicas across all bidders
must outweigh the total decrease caused by unmatching surrogates.
We therefore get:

\[
\E[\frac{\epsilon}{r}\sum_{j}\sum_{s\in U_{j}}p_{j}(s)-\sum_{s\in T_{j}}p_{j}(s)]\leq\E[\frac{1}{r}\sum_{j}\sum_{r_{j}}\gamma r_{j}(\phi_{j}(s_{j}))+\gamma r_{j}(\phi_{j}(s_{j}'))+\delta_{j}(\phi_{j}(s_{j}'))+\eta(\phi_j(s_j))]
\]

Consider now the following allocation algorithm (\emph{not} a mechanism, and certainly not truthful).
When bidder $j$ reports $r_{j}$, the algorithm selects the allocation
output by $M$ on input $(s_{1},\ldots,s_{m})$. This is clearly a
feasible allocation algorithm, and it's also clear that the term $\E[\frac{1}{r}\sum_{j}\sum_{r_{j}}r_{j}(\phi_{j}(s_{j}))]$
exactly computes the expected welfare achieved by this algorithm when
the type of buyer $j$ is drawn from $D_{j}$. We can similarly define
an allocation algorithm that replaces $r_{j}$ with $s_{j}'$, and
one that samples $\delta_{j}\leftarrow D_{j}^{+}-D_{j}$ and replaces
$\delta_{j}$ with $s_{j}'$ (and ditto for $\eta$). Now that we have concrete allocation
algorithms that match the terms on the right-hand side exactly, we
can bound it as:

\[
\E[\frac{1}{r}\sum_{j}\sum_{r_{j}}\gamma r_{j}(\phi_{j}(s_{j}))+\gamma r_{j}(\phi_{j}(s_{j}'))+\delta_{j}(\phi_{j}(s_{j}'))]\leq2\gamma\val(D)+\val(\delta) + \val(\eta)
\]

Observe, however, that in $\eta(\cdot)$ every bidder values every item at $\eta$, so $\val(\eta) = \eta \cdot n$ ($n$ is the number of items). Lastly, because this claim holds for all $\gamma>0$, we may let $\gamma\rightarrow0$
and the right-hand bound becomes just $\val(\delta)+\eta n$. Some care must
be taken if $\val(D)=\infty$, but similar tricks to those used in
the proof of Lemma~\ref{lem:4} suffice. Essentially, one can take
an increasing limit of truncations of $D$, calling them $D^{C}$
(the same distribution $D$ but replacing $v(\cdot)$ with the $0$-function
if $v([n])>C$). Clearly all $\val(D^{C})$ is finite for all $C$,
and clearly $\lim_{C\rightarrow\infty}\rev^{M}(D^{C})=\rev^{M}(D)$
for any mechanism $M$. So one can use the above analysis on the bounded
distributions $D^{C}$ and take a limit.
\end{proof}
With the lemma above, our proof is complete. There is a high-cardinality
matching $W$ that provides revenue exactly $\rev^{M}(D)$, if only
it were chosen by VCG. But VCG may choose a different matching, and
we may lose some revenue. The lemma bounds how much revenue can be
lost, and provides the bound in the theorem.


\begin{ack}
We would like to thank Moshe Babaioff, Hu Fu, Nicole Immorlica, and
Brendan Lucier for numerous suggestions and helpful discussions.
\end{ack}

\bibliographystyle{acmsmall}
\bibliography{bib}

\ifFULL
\appendix

\fi
\end{document}